\documentclass[twocolumn, superscriptaddress,aps, showpacs, altaffiliationletter]{revtex4}

\usepackage{amsfonts}
\usepackage{amsmath}
\usepackage{epsfig}
\usepackage{bm}

\usepackage{amssymb}
\input{epsf}

\begin{document}

\title{Channeling chaotic transport in a wave-particle experiment}

\author{A. Macor}
\email[e-mail address:]{macor@up.univ-mrs.fr} \affiliation{Equipe
turbulence plasma, CNRS,  Centre Saint-J\'er\^ome, case 321, 13397
Marseille cedex 20, France}

\author{F. Doveil}
\affiliation{Equipe turbulence plasma, CNRS,  Centre
Saint-J\'er\^ome, case 321, 13397 Marseille cedex 20, France}

\author{C. Chandre}
 \affiliation{Centre de
Physique Th\'{e}orique \footnote{Unit\'{e} Mixte de Recherche (UMR
6207) du CNRS, et des universit\'{e}s Aix-Marseille I, Aix-Marseille
II et du Sud Toulon-Var. Laboratoire affili\'{e} \`{a} la FRUMAM (FR
2291). Laboratoire de Recherche Conventionn\'e du CEA (DSM-06-35)},
CNRS, Luminy, case 907, F-13288 Marseille, France.}

\author{G. Ciraolo}
\affiliation{Centre de Physique Th\'{e}orique \footnote{Unit\'{e}
Mixte de Recherche (UMR 6207) du CNRS, et des universit\'{e}s
Aix-Marseille I, Aix-Marseille II et du Sud Toulon-Var. Laboratoire
affili\'{e} \`{a} la FRUMAM (FR 2291). Laboratoire de Recherche
Conventionn\'e du CEA (DSM-06-35)}, CNRS, Luminy, case 907, F-13288
Marseille, France.}

\affiliation{ Association Euratom-CEA, DRFC/DSM/CEA, CEA Cadarache,
F-13108 St. Paul-lez-Durance, France}

\author{R.Lima}
\affiliation{Centre de Physique Th\'{e}orique \footnote{Unit\'{e}
Mixte de Recherche (UMR 6207) du CNRS, et des universit\'{e}s
Aix-Marseille I, Aix-Marseille II et du Sud Toulon-Var. Laboratoire
affili\'{e} \`{a} la FRUMAM (FR 2291). Laboratoire de Recherche
Conventionn\'e du CEA (DSM-06-35)}, CNRS, Luminy, case 907, F-13288
Marseille, France.}

\author{M. Vittot}
\affiliation{Centre de Physique Th\'{e}orique \footnote{Unit\'{e}
Mixte de Recherche (UMR 6207) du CNRS, et des universit\'{e}s
Aix-Marseille I, Aix-Marseille II et du Sud Toulon-Var. Laboratoire
affili\'{e} \`{a} la FRUMAM (FR 2291). Laboratoire de Recherche
Conventionn\'e du CEA (DSM-06-35)}, CNRS, Luminy, case 907, F-13288
Marseille, France.}

%
%
\begin{abstract}
\texttt{[A numerical and experimental study of a control method
aimed at channeling chaos by building barriers in phase space is
performed on a paradigm for wave-particle interaction, i.e., a
traveling wave tube. Control of chaotic diffusion is achieved by
adding small apt modifications to the system with a low additional
cost of energy. This modification is realized experimentally through
additional waves with small amplitudes. Robustness of the method is
investigated both numerically and experimentally.]}
\end{abstract}

\maketitle
\section{Introduction}
The interaction of a beam of charged particles with
electromagnetic waves is ubiquitous in physics, and it is central
to many useful devices such as particle accelerators, plasma
fusion experiments or free electron lasers. In these experimental
set-ups, the waves are used to accelerate the particles or to
guide them by assigning a specific mean trajectory. However, the
dynamics of these systems is usually characterized by the
competition of many degrees of freedom and thus, shows generically
chaotic behaviors. Such behaviors represent a limit to the
achievement of high performances in these devices. Consequently,
there has been a growing interest in controlling chaos which here
means to reduce it when and where it is undesirable and to
increase it when it is useful.

The sensitivity of chaotic systems to small perturbations
triggered a strong interdisciplinary effort to control
chaos~\cite{c1,c2,c3,c4,c5,c6,c7,c8}. After the seminal
work on optimal control by Pontryagin~\cite{c9}, efficient methods
were proposed for controlling chaotic systems by nudging targeted
trajectories~\cite{c10,c11,c12,c13,c14}. However, for many
body experiments such methods are hopeless due to the high number
of trajectories to deal with simultaneously.

It was recently proposed a local control method~\cite{chan06}
which aims at building barriers in phase space and hence confines
all the trajectories rather than following them individually.
These barriers are stable structures in phase space like for
instance invariant tori, which are generically broken by the
perturbation. The reduction of chaotic behaviors is achieved by
using a small apt perturbation of the system which keeps its
Hamiltonian structure.

In this article, we consider a traveling wave tube (TWT) specially
designed to study wave particle interaction which is used to
investigate experimentally the control method and its robustness.
The dynamics in this experimental apparatus can be accurately
represented using a Hamiltonian which describes the motion of a
charged particle (with unit mass) interacting with two
electrostatic waves~\cite{c16}~:

\begin{eqnarray}
\label{eqn:H}
    H(p,x,t)&=&\frac{p^2}{2}+\varepsilon_1\cos (k_1 x-\omega_1
    t+\varphi_1)\nonumber
    \\[0.5\baselineskip]
    &&+\varepsilon_2\cos (k_2 x-\omega_2
    t+\varphi_2),
    \label{Ham1}
\end{eqnarray}
where $(p,x)\in {\mathbb R}\times [0,L]$ are the momentum and
position of the particle in a tube of length $L$. The amplitudes,
wave numbers, frequencies and phases of the two waves are denoted
respectively $\varepsilon_i$, $k_i$, $\omega_i$ and $\varphi_i$
for $i=1,2$. We notice that the beam intensity is sufficiently low
such that the wave growth rate is negligible upon the length of
the experiment that is we are in the test-particle regime.

Generically, the dynamics of the particles governed by
Hamiltonian~(\ref{Ham1}) is a mixture of regular and chaotic
behaviors, mainly depending on the amplitudes of the waves. The
Chirikov parameter~\cite{Chirikov} defined as the ratio between
the two half-width of the primary resonances by the distance
between  these resonances, i.e.,
\begin{equation}
\label{eqn:pCh}
    s=\frac{{2}(\sqrt{{|\varepsilon_1}|}+\sqrt{{|\varepsilon_2|}})}{|{\omega_2}/{k_2}-{\omega_1}/{k_1}|},
\end{equation}
gives a first rough approximation of the chaoticity degree of the
system. Hamiltonian~(\ref{eqn:H}) has a typical behavior of
integrable system for small values of this parameter ($s$$\ll1$).
For large enough amplitudes of the waves ($s\sim 1$), large scale
chaos occurs in phase space. As a consequence, the particle can
have an arbitrary velocity in between the two phase velocities of
the waves ($\omega_2/k_2$ and $\omega_1/k_1$). In this TWT, such
typical chaotic behavior has been observed directly~\cite{Macor1}.
This chaotic diffusion of the particles in phase space can be
reduced by using an apt control term which consists here as an
additional wave (or more generally a set of waves) of small
amplitude. The characteristics of this additional wave are
computed explicitly, and then the wave is injected in addition to
the two others. The results presented in this article were
announced in Ref.~\cite{c18}.

The paper is organized as follows~: The control method is briefly
recalled in Sec.~\ref{loc:method} and its application to the
considered Hamiltonian is presented in Sec.~\ref{loc:computation}.
Numerical investigations of the effect of the control term and its
robustness are reported in Secs.~\ref{loc:simulation}
and~\ref{loc:robust}. In Sec.~\ref{experiment:setup}, a
description of the experimental set-up precedes the results of the
implementation of the control term shown in
Sec.~\ref{experiment:implementation} as well as its robustness in
Sec.~\ref{experiment:robust}.

\section{Local control method applied to a two wave model}
\label{loc:control} The Hamiltonian of an integrable system can be
locally written as a function $H_0({\bf A})$ of the action
variables ${\bf A}=(A_1,A_2,\ldots,A_d)\in{\mathbb R}^d$, i.e.\ it
does not depend on the conjugate angle variables
${\bm\theta}=(\theta_1,\theta_2,\ldots,\theta_d)\in{\mathbb T}^d
={\mathbb R}^d/(2\pi {\mathbb Z})^d$, where ${\mathbb T}^d$ is the
 $d$-dimensional torus parameterized by $[0,2\pi[^d$.
The equations of motion for $H_0({\bf A})$ show that the action
variables are constant, and consequently the trajectories with
given actions $\bf{A}$ are confined to evolve on a $d$-dimensional
torus with frequency vector ${\bm \omega}_0({\bf A})= \partial
H_0/\partial {\bf A}$. The dynamics on this torus is periodic or
quasi-periodic: ${\bm \theta}(t)={\bm \omega}_0({\bf A}) t+{\bm
\theta}(0)$ with frequency vector ${\bm \omega}_0({\bf A})$. In
the particular case given by Hamiltonian~(\ref{Ham1}) an
integrable situation is given by $\varepsilon_1=\varepsilon_2=0$
so that the dynamics of the integrable system  $H=p^2/2$ is
characterized by constant velocity ($p=const$). A monokinetic beam
of charged particles remains monokinetic.

If the system described by $H_0$ is perturbed, i.e.\ we consider
the Hamiltonian
\begin{equation}
\label{eqn:Hgene} H({\bf A},{\bm \theta})=H_0({\bf A})+V({\bf
A},{\bm \theta})\nonumber,
\end{equation}
the integrability is generically lost and the system becomes
chaotic. Even if KAM theorem establishes the stability with
respect to small perturbations of invariant tori with a
sufficiently incommensurate frequency vector these tori are
destroyed when the amplitude of the perturbation $V$ is large
enough. The break-up of invariant tori leads to a loss of
stability of the system until the last invariant torus of the
integrable case is destroyed and then large scale diffusion occurs
in phase space. In the case of a beam of charged particles whose
dynamics is given by Hamiltonian~(\ref{Ham1}), for
$\varepsilon_1$, $\varepsilon_2$ sufficiently large, an initially
monokinetic beam will spread in velocity due to this diffusion.

\subsection{Expression of the local control term}
\label{loc:method} The aim is to provide an explicit expression
for an additional perturbation such that a specific invariant
torus is reconstructed in the modified system. We state here the
main result which has been extensively described in
Ref.~\cite{chan06}~: We consider Hamiltonian systems written as
$$
H({\bf A},{\bm\theta})={\bm \omega}\cdot {\bf A}+ W({\bf
A},{\bm\theta}),$$ where $\bm\omega$ is a non-resonant vector of
${\mathbb R}^d$. Without loss of generality, we consider a region
near ${\bf A}={\bf 0}$ (by translation of the actions) and, since
the Hamiltonian is nearly integrable, the perturbation $W$ has
constant and linear parts in actions of order $\varepsilon$, i.e.\
\begin{equation}
\label{eqn:e4V} W({\bf A},{\bm\theta})=\varepsilon
v({\bm\theta})+\varepsilon {\bf w}({\bm\theta})\cdot {\bf
A}+Q({\bf A},{\bm\theta}),
\end{equation}
where $Q$ is of order $O(\Vert {\bf A}\Vert ^2)$. We notice that
for $\varepsilon=0$, the Hamiltonian $H$ has an invariant torus
with frequency vector ${\bm\omega}$ at ${\bf A}={\bf 0}$ for any
$Q$ not necessarily small. The controlled Hamiltonian we construct
is
\begin{equation}
\label{eqn:gene} H_c({\bf A},{\bm\theta})={\bm \omega}\cdot {\bf
A}+ W({\bf A},{\bm\theta})+ f({\bm \theta}).
\end{equation}
The control term $f$ we construct only depends on the angle
variables and is given by
\begin{equation}
\label{eqn:exf} f({\bm\theta})=W({\bf 0},{\bm\theta})-
W\left(-\Gamma \partial_{\bm\theta} W({\bf
0},{\bm\theta}),{\bm\theta}\right),
\end{equation}
where $\partial_{\bm\theta}$ is the derivative operator with
respect to $\bm\theta$, and $\Gamma$ is a linear operator defined
as a pseudo-inverse of ${\bm\omega}\cdot
\partial_{\bm\theta}$, i.e.\ acting on $W=\sum_{{\bf k}}W_{{\bf
k}} {\mathrm e}^{i{\bf k}\cdot{\bm\theta}}$ as
$$
\Gamma W=\sum_{{\bm\omega}\cdot{\bf k}\not= 0} \frac{W_{{\bf
k}}}{i{\bm\omega}\cdot{\bf k}} {\mathrm e}^{i{\bf
k}\cdot{\bm\theta}}.
$$
Note that $f$ is of order $\varepsilon^2$. For any function $W$,
Hamiltonian~(\ref{eqn:gene}) has an invariant torus with frequency
vector close to ${\bm\omega}$. The equation of the torus which is
restored by the addition of $f$ is
\begin{equation}
\label{eqn:eto} {\bf A}=-\Gamma \partial_{\bm\theta} W({\bf
 0},{\bm\theta}),
\end{equation}
which is of order $\varepsilon$ for $W$ given by
Eq.~(\ref{eqn:e4V})

\begin{figure}[tbp]
\includegraphics[width=7.5cm,height=7.2cm]{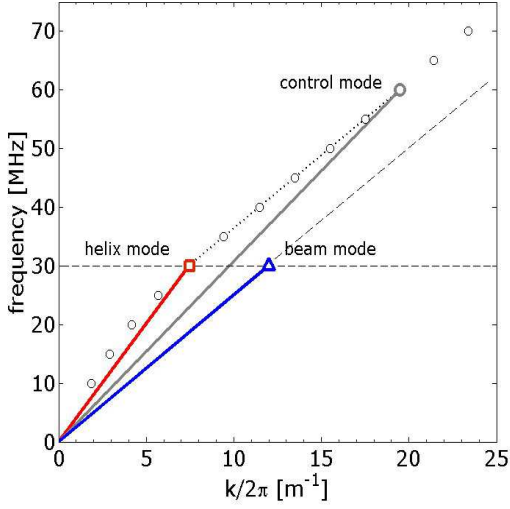}
\caption{TWT dispersion relation (circles) with the helix mode at
30{\,\rm MHz} (square) and the beam mode at the same frequency but
with propagation velocity choose equal to about $2.5\times
10^6${\,\rm m/s} (triangle); the control wave corresponds to the
beating of these two modes
 } \label{fig:reldisp}
\end{figure}

\subsection{Computation of the control term for a two wave system}
\label{loc:computation} We consider Hamiltonian~(\ref{Ham1}) with
two waves, where the wavenumbers are chosen according to a
dispersion relation $k_1=K(\omega_1)$ and $k_2=K(\omega_2)$
plotted on Fig.~\ref{fig:reldisp}.

In order to compute $f$, Hamiltonian (\ref{Ham1}) with 1.5 degrees
of freedom is mapped into an autonomous Hamiltonian with two
degrees of freedom by considering that $t \mbox{ mod }2\pi$ is an
additional angle variable. We denote $E$ its conjugate action. The
autonomous Hamiltonian is
\begin{eqnarray}
\label{Htwt_autonoma}
H(E,p,x,t)&=&E+\frac{p^2}{2}+\varepsilon_1\cos (k_1x-\omega_1
t+\varphi_1)\nonumber
\\[0.5\baselineskip]
&&+\varepsilon_2\cos (k_2 x-\omega_2 t+\varphi_2).
\end{eqnarray}
Then, the momentum $p$ is shifted by $\omega$ in order to define a
local control in the region $p\approx 0$. The Hamiltonian is
rewritten as

\begin{eqnarray}
H&=&E+\omega p+\varepsilon_1 \cos (k_1 x-\omega_1
t+\varphi_1)\nonumber \nonumber
    \\[0.5\baselineskip]
    &&+\varepsilon_2\cos (k_2 x-\omega_2
t+\varphi_2)+\frac{p^2}{2}. \label{Htwt_pshift}
\end{eqnarray}

We rewrite Hamiltonian~(\ref{Htwt_pshift}) into the
form~(\ref{eqn:e4V})~ where:
\begin{eqnarray*}
&& \varepsilon v(x,t)=\varepsilon_1\cos (k_1 x-\omega_1
t+\varphi_1)+\varepsilon_2\cos (k_2 x-\omega_2 t+\varphi_2),\\
&& w(x,t)=0,\\
&& Q(p,x,t)=p^2/2.
\end{eqnarray*}
The frequency vector of the selected invariant torus is
${\bm\omega}=(\omega,1)$. From Eq.~(\ref{eqn:exf}) we have that
$f$ is given by
$$
f(x,t)=-\frac{\varepsilon^2}{2} (\Gamma \partial_{x} v) ^2,
$$
which is
\begin{eqnarray}
    &f(x,t)= &-\frac{1}{2}\big[ \frac{\varepsilon_1 k_1}{\omega
k_1-\omega_1}\cos (k_1x-\omega_1t+\varphi_1)\nonumber\\
& &+\frac{\varepsilon_2 k_2}{\omega
k_2-\omega_2}\cos(k_2x-\omega_2t+\varphi_2)\big]^2,
    \label{eqn:fTWT}
\end{eqnarray}
provided $\omega\neq\omega_1/k_1$ and  $\omega\neq\omega_2/k_2$.
Adding this exact control term to Hamiltonian~(\ref{Ham1}), the
following invariant rotational torus is restored~:
\begin{eqnarray}
p(x,t)&=&\omega-\frac{\varepsilon_1 k_1}{\omega k_1-\omega_1} \cos
(k_1x-\omega_1t+\varphi_1) \nonumber
    \\[0.5\baselineskip]
&&-\frac{\varepsilon_2 k_2}{\omega
k_2-\omega_2}\cos(k_2x-\omega_2t+\varphi_2).
\end{eqnarray}
This barrier of diffusion prevents a beam of particles to diffuse
everywhere in phase space. We emphasize that the barrier persists
for all the magnitudes of the waves
$(\varepsilon_1,\varepsilon_2)$.

The control term~(\ref{eqn:fTWT}) has four Fourier modes,
$(2k_1,-2\omega_1)$, $(2k_2,-2\omega_2)$,
$((k_1+k_2),-(\omega_1+\omega_2))$ and
$((k_1-k_2),-(\omega_1-\omega_2))$. If we want to restore an
invariant torus in between the two primary resonances
approximately located at $p\approx \omega_1/k_1$ and $p\approx
\omega_2/k_2$, the frequency $\omega$ has to be chosen between the
two group velocities of the waves. If we consider a beam of
particles with a velocity in between the velocities of the waves,
i.e., $v_1=\omega_1/k_1$ and $v_2=\omega_2/k_2$, the main Fourier
mode of the control term is
\begin{eqnarray*}
f_2&=&-\frac{\varepsilon_1\varepsilon_2 k_1 k_2}{2(\omega
k_1-\omega_1)(\omega k_2-\omega_2)} \nonumber
    \\[0.5\baselineskip]
&&\times \cos
[(k_1+k_2)x-(\omega_1+\omega_2)t+\varphi_1+\varphi_2].
\end{eqnarray*}

A convenient choice is $\omega=(v_1+v_2)/2$ and the control term
is given by:
\begin{equation}
f_2=\frac{2\varepsilon_1\varepsilon_2}{(v_1-v_2)^2}\cos
[(k_1+k_2)x-(\omega_1+\omega_2)t+\varphi_1+\varphi_2].\label{eqn:f2app}
\end{equation}
Using this approximate control term does not guarantee the
existence of an invariant torus. However, since the difference
between $f$ given by Eq.~(\ref{eqn:fTWT}) and $f_2$ given by
Eq.~(\ref{eqn:f2app}) is small, it is expected that for a Chirikov
parameter $s$ not too large, the effect of the control term is
still effective and the barrier is restored close to
\begin{eqnarray}
p(x,t)&=&\frac{v_1+v_2}{2}+\frac{2\varepsilon_1\cos(k_1x-\omega_1t+\varphi_1)}
{v_1-v_2} \nonumber
    \\[0.5\baselineskip]
&&-\frac{2\varepsilon_2\cos(k_2x-\omega_2t+\varphi_2)}{v_1-v_2}.
\end{eqnarray}

\subsection{Numerical results}
\label{loc:simulation} 
\begin{figure}[tbp]
\includegraphics[width=7.5cm,height=7.2cm]{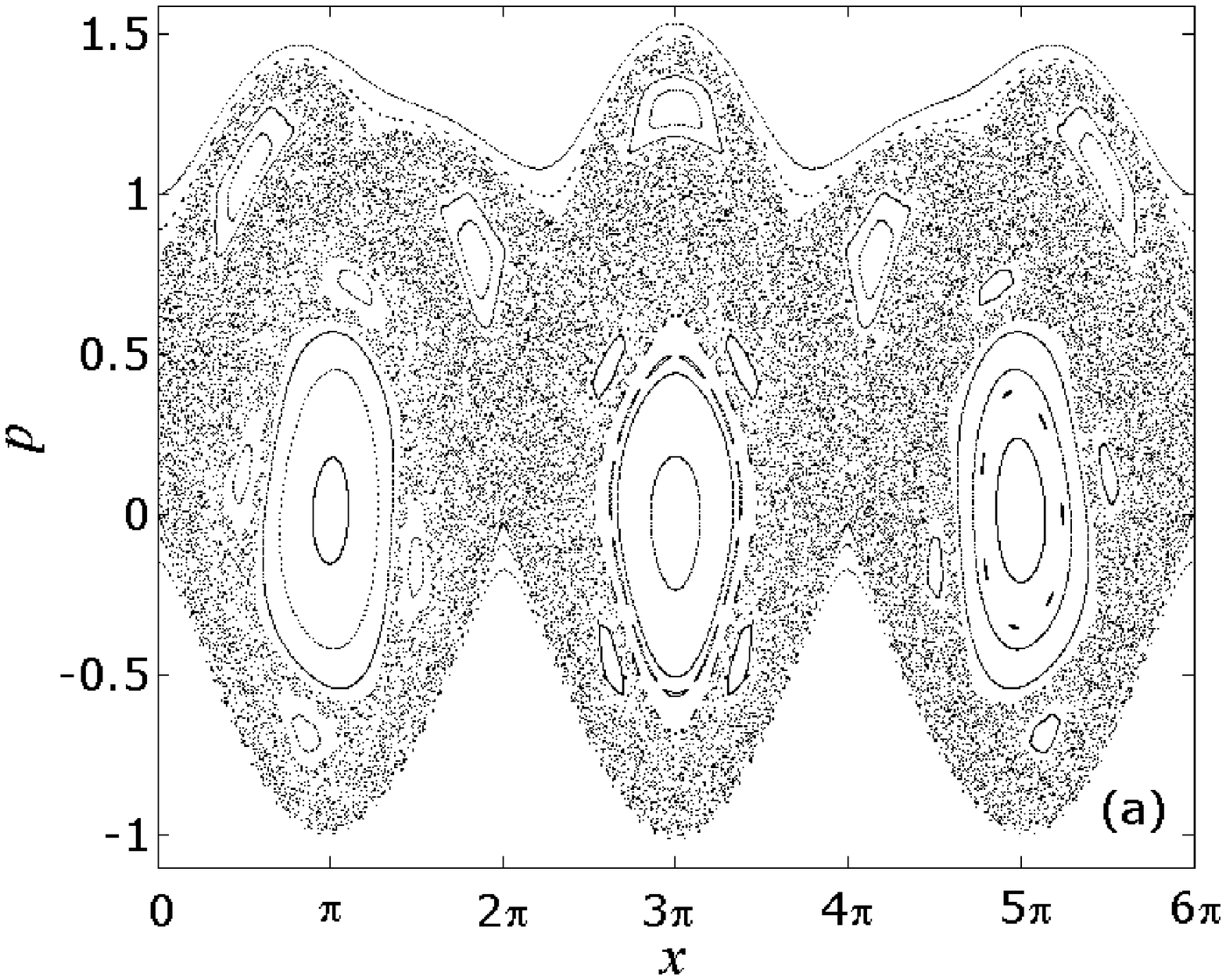}
\includegraphics[width=7.5cm,height=7.2cm]{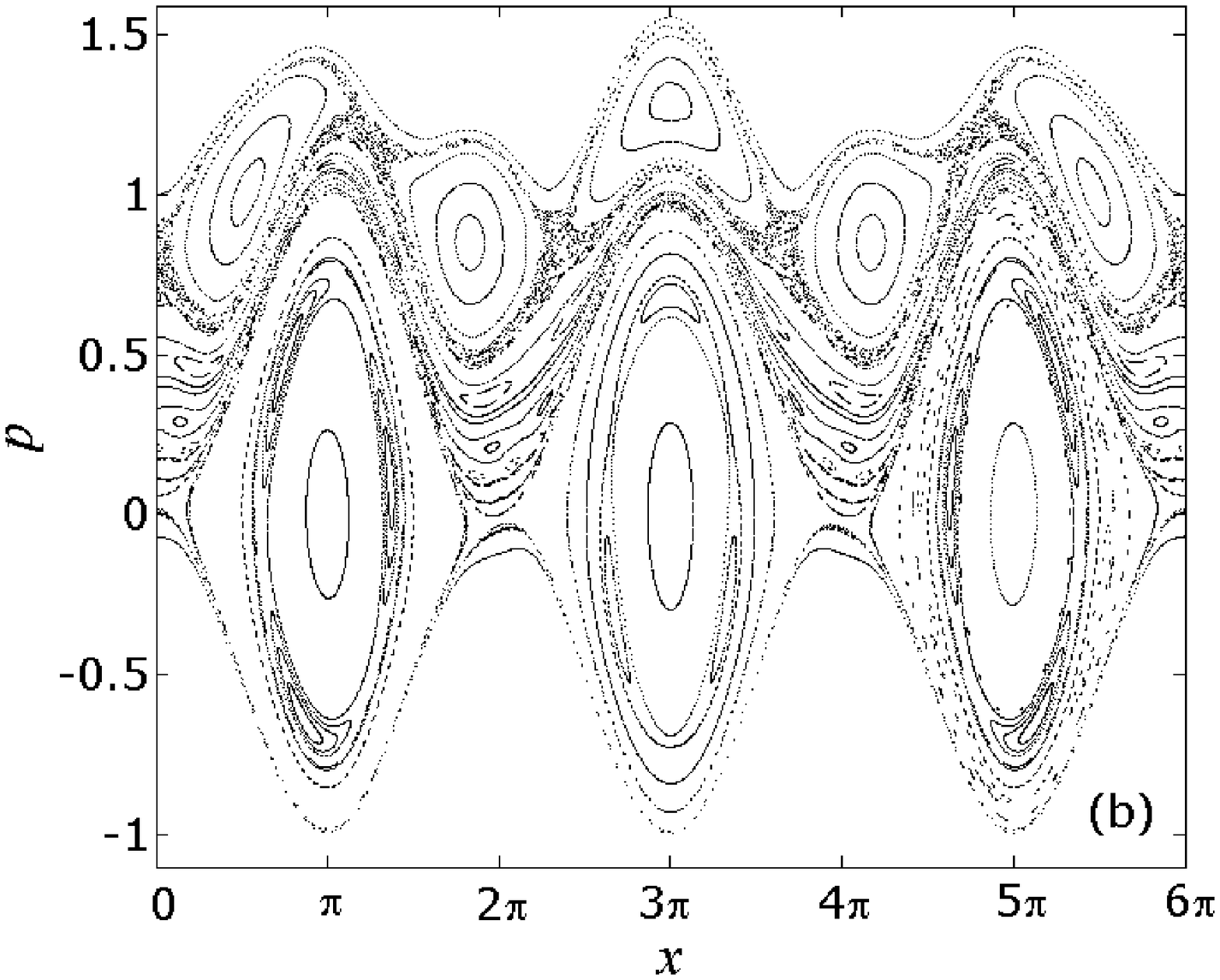}
\includegraphics[width=7.5cm,height=7.2cm]{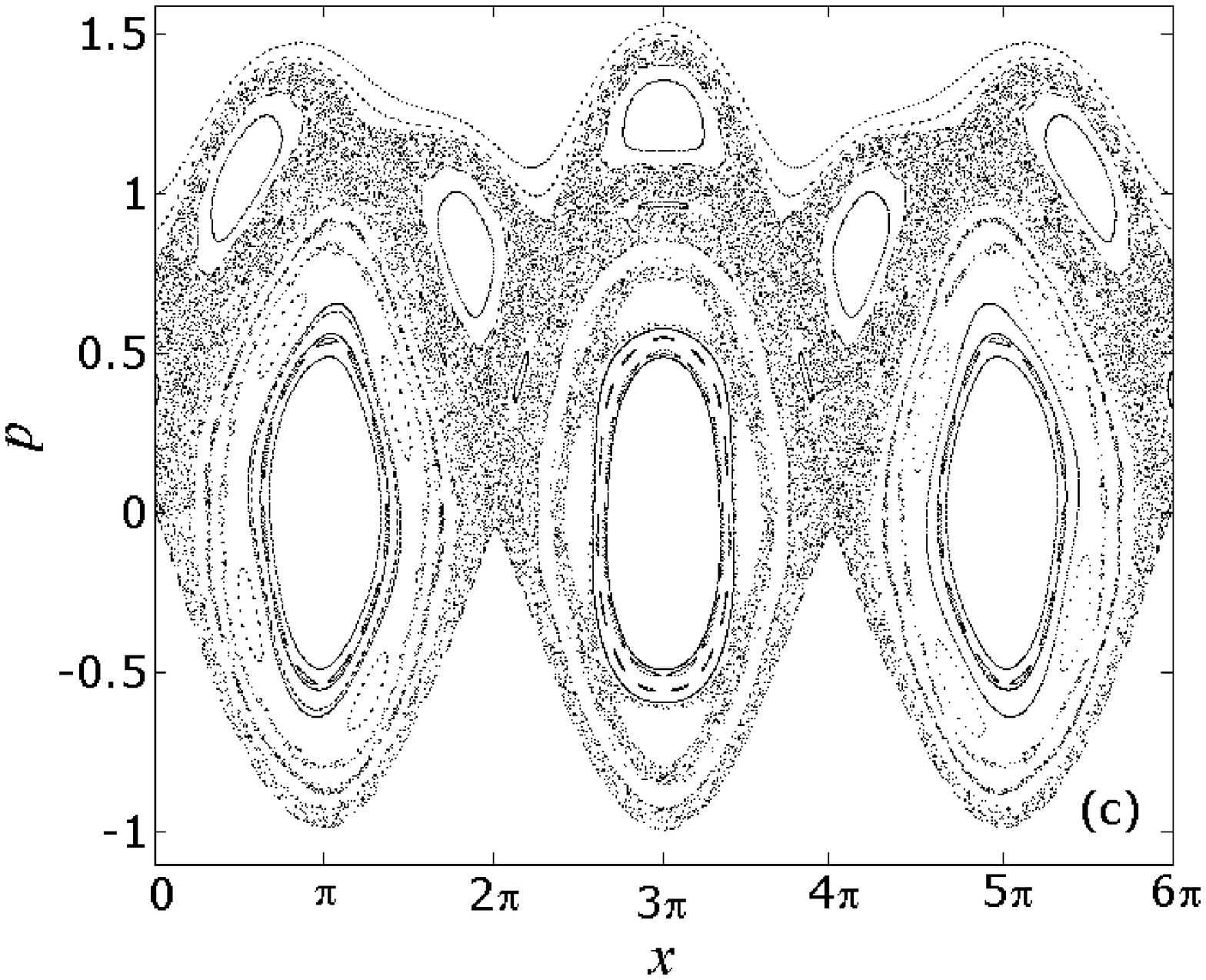}
\caption{Poincar\'e sections of Hamiltonian (\ref{Ham:sim}) for
$s=1.27$ without control term (panel (a)), plus the exact control
term~(\ref{eqn:fsimTWT}) (panel (b)) and plus the approximate
control term~(\ref{eqn:fsimappTWT}) (panel (c)).}
\label{fig:Poinc2}
\end{figure}

In this section we perform a numerical
investigation of the effect of the exact and approximate control
terms on the electron beam dynamics. We introduce the parameter
$r$ given by the ratio of the two wave amplitudes
$r=\varepsilon_1/\varepsilon_2$. In order to reproduce as close as
possible the experimental set-up described in the next section
(see also~\cite{Macor1}), we consider the following values of
amplitudes, wave numbers, frequencies and phases of the two
electrostatic waves:
$(\varepsilon_1,k_1,\omega_1,\varphi_1)=(\varepsilon r,1,0,0)$ and
$(\varepsilon_2,k_2,\omega_2,\varphi_2)=(\varepsilon,k,k,0)$. Thus
Hamiltonian~(\ref{Ham1}) can be written as
\begin{equation}
H(p,x,t)=\frac{p^2}{2}+\varepsilon r\cos x+\varepsilon \cos
[k(x-t)], \label{Ham:sim}
\end{equation}
i.e. $v_1=0$ and $v_2=1$. We perform simulations with $r=0.082$
and $k=5/3$. The amplitudes of the waves are determined by $r$ and
$\varepsilon$ that are related to the Chirikov parameter by the
following equation:
\begin{equation}
s=2\sqrt{\varepsilon}(\sqrt{r}+1).
\end{equation}
The value of $\varepsilon$ will be given by $s$. In the following
we consider two values of $s$, that is $s=0.85$
($\varepsilon\sim0.11$) and $s=1.27$ ($\varepsilon\sim0.24$). In
this case the expression of the exact control term given by
Eq.~(\ref{eqn:fTWT}) becomes
\begin{equation}
  f(x,t)= -2\varepsilon^2(r\cos x-\cos k(x-t))^2,
    \label{eqn:fsimTWT}
\end{equation}
while the approximate control term given by Eq.~(\ref{eqn:f2app})
is
\begin{equation}
  f(x,t)= 2\varepsilon^2r\cos[(k+1)x-kt].
    \label{eqn:fsimappTWT}
\end{equation}
\begin{figure}[tbp]
\includegraphics[width=7.5cm,height=7.2cm]{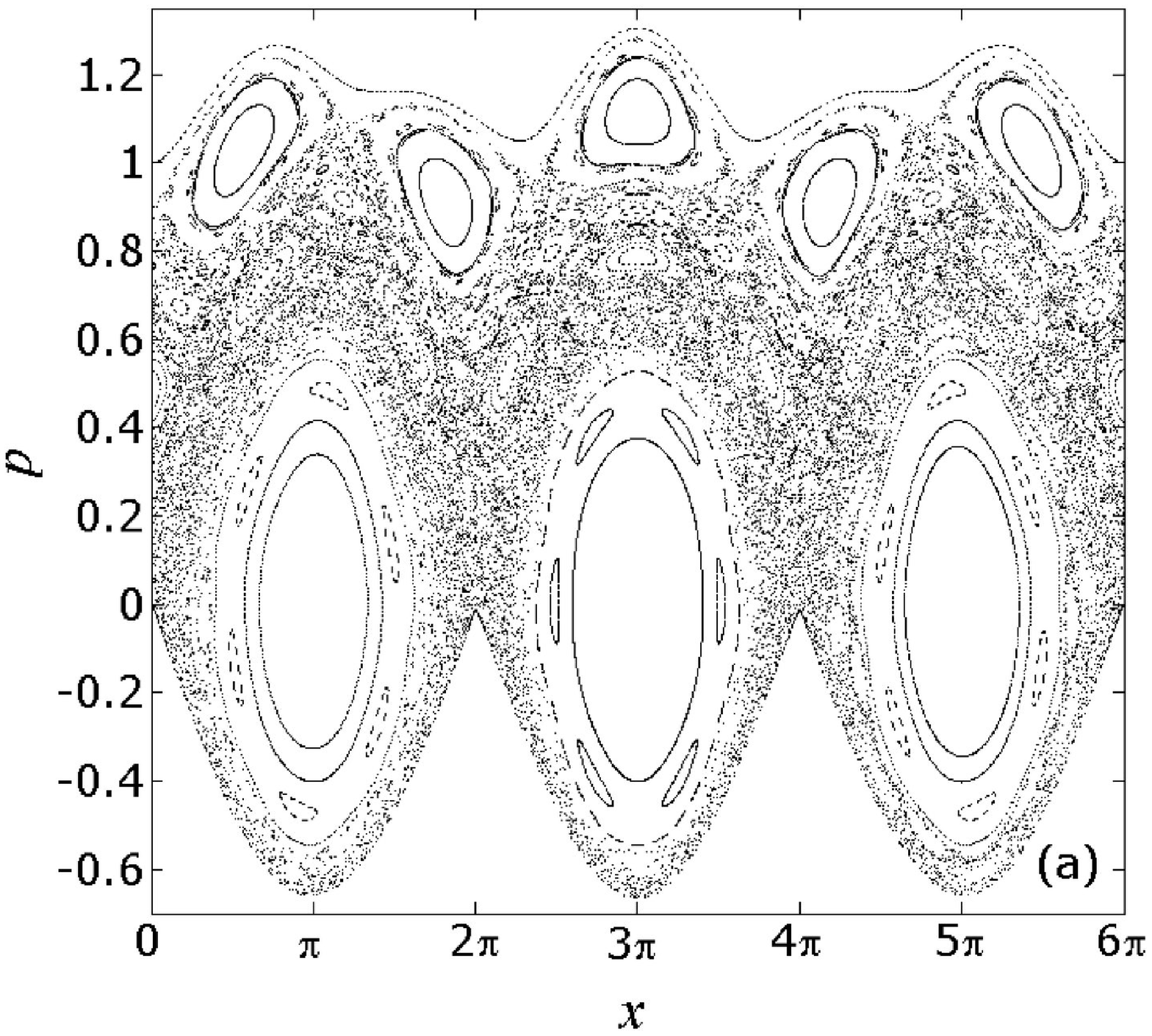}
\includegraphics[width=7.5cm,height=7.2cm]{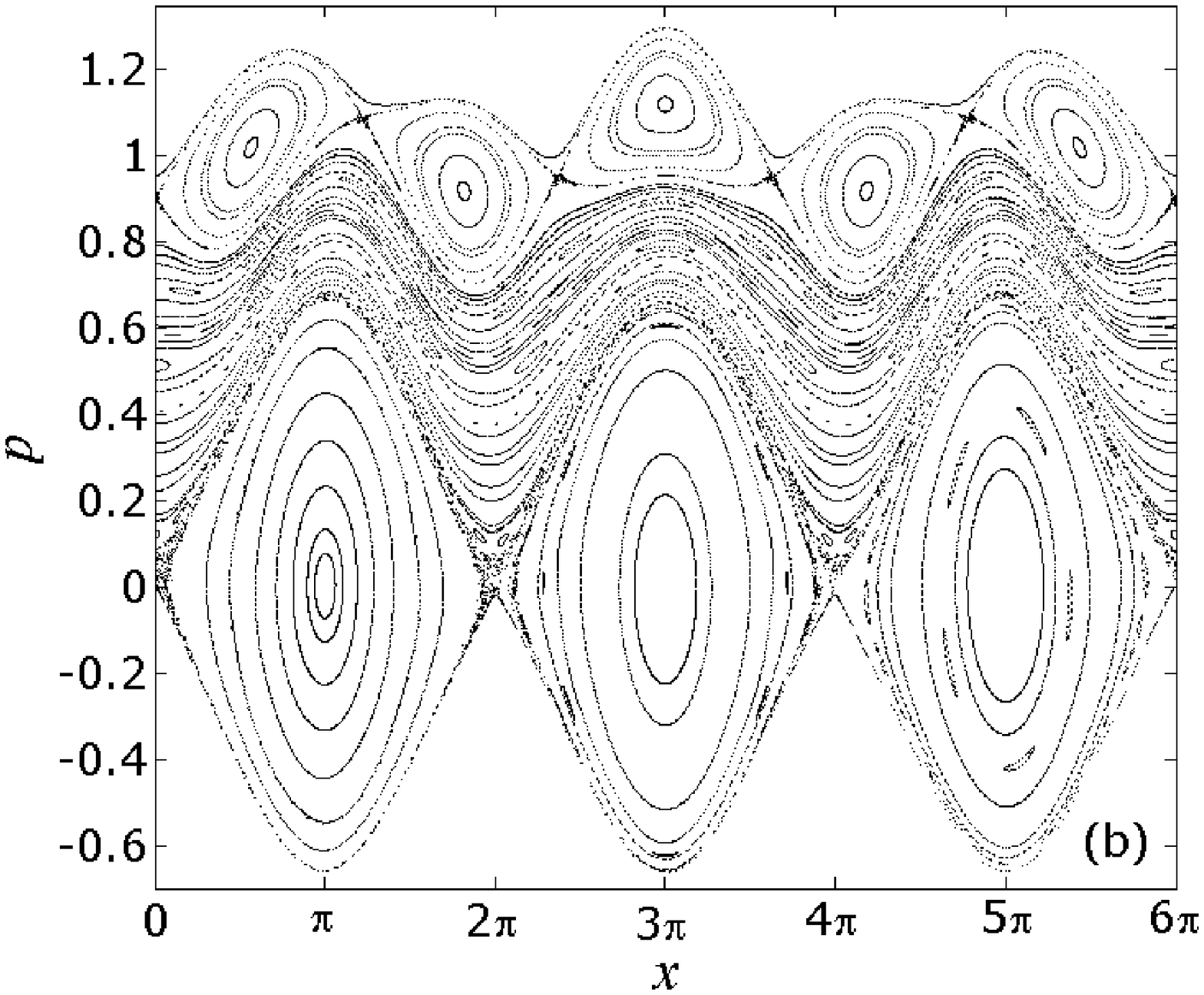}
\includegraphics[width=7.5cm,height=7.2cm]{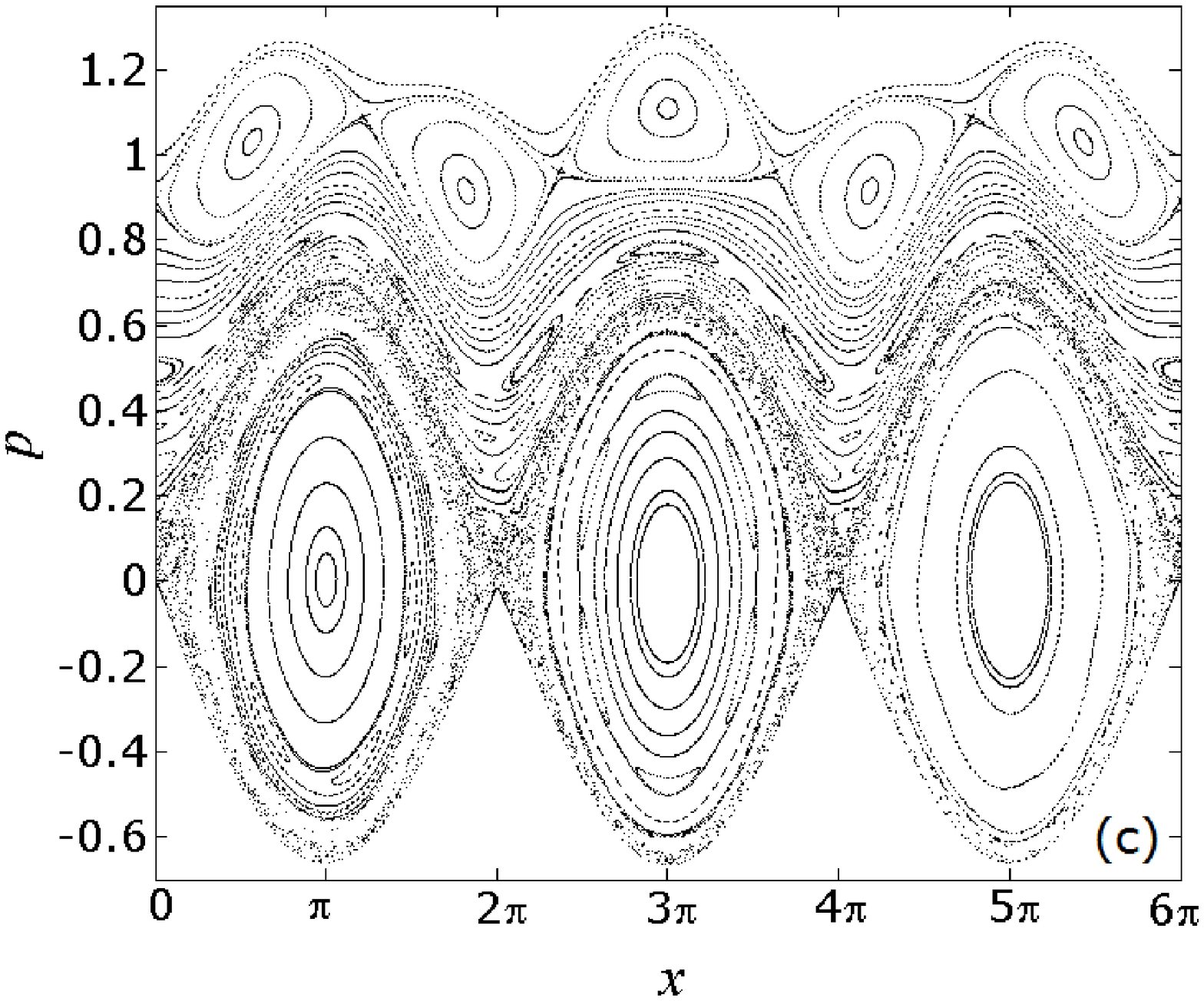}
\caption{Poincar\'e sections of Hamiltonian (\ref{Ham:sim}) for
$s=0.85$ without control term (panel (a)), plus the exact control
term~(\ref{eqn:fsimTWT}) (panel (b)) and plus the approximate
control term ~(\ref{eqn:fsimappTWT})(panel (c)).}
\label{fig:Poinc1}
\end{figure}
Poincar\'e sections of Hamiltonian~(\ref{Ham:sim}) computed for
two values of the Chirikov parameter, $s=0.85$ and $s=1.27$, are
depicted in Figs.~\ref{fig:Poinc1}-\ref{fig:Poinc2}, panels (a).
We notice that in both cases no rotational invariant tori survive
and therefore trajectories can diffuse over the whole phase space
in between the two primary resonances.
As expected, when the exact control term given by
Eq.~(\ref{eqn:fsimTWT}) is added to the original Hamiltonian, then
rotational invariant tori are restored. This is shown by the two
Poincar\'e sections in Figs.~\ref{fig:Poinc1} and
~\ref{fig:Poinc2}, panels (b), corresponding to the two values
$s=0.85$ and $s=1.27$ of the Chirikov parameter. We notice that
$s=1.27$ corresponds to a chaotic regime where the two resonances
overlap according to Chirikov criterion ($s\ge 1$). Nevertheless
the exact control term is able to reconstruct the invariant torus
predicted by the method and to regularize a quite large region
around the recreated invariant torus.

In order to study the effect of a simplified control term on the
electron beam dynamics we perform numerical simulations adding the
control term given by Eq.~(\ref{eqn:fsimappTWT}) to Hamiltonian
(\ref{Ham:sim}). As one can see from the Poincar\'e section
depicted in Fig.~\ref{fig:Poinc1}, panels (c), the effect of the
approximate control term is still present with the recreation of a
set of invariant tori for $s=0.85$. However, this regularization
apparently disappears when we consider the fully chaotic regime
with $s=1.27$ (see Fig.~\ref{fig:Poinc2}, panel (c)). Nevertheless
the approximate control term has still a significant effect on the
reduction of chaotic diffusion. This fact can be observed on the
probability distribution functions of the electron beam velocity.
This diagnostic will also be used in the experiment in order to
see the effect of the control terms.

\begin{figure}
\includegraphics[width=8.0cm, height=7.0cm]{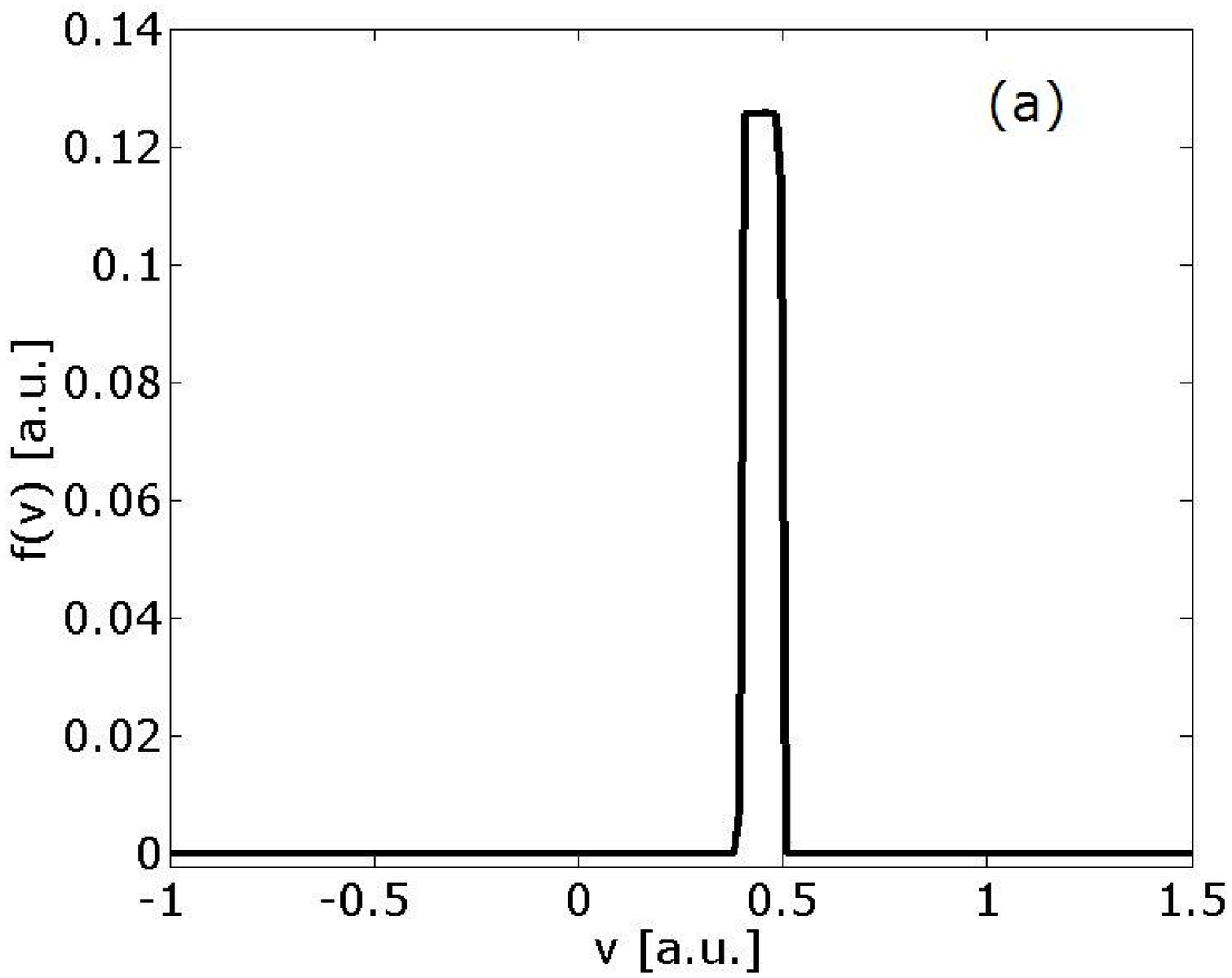}
\includegraphics[width=8.0cm, height=7.0cm]{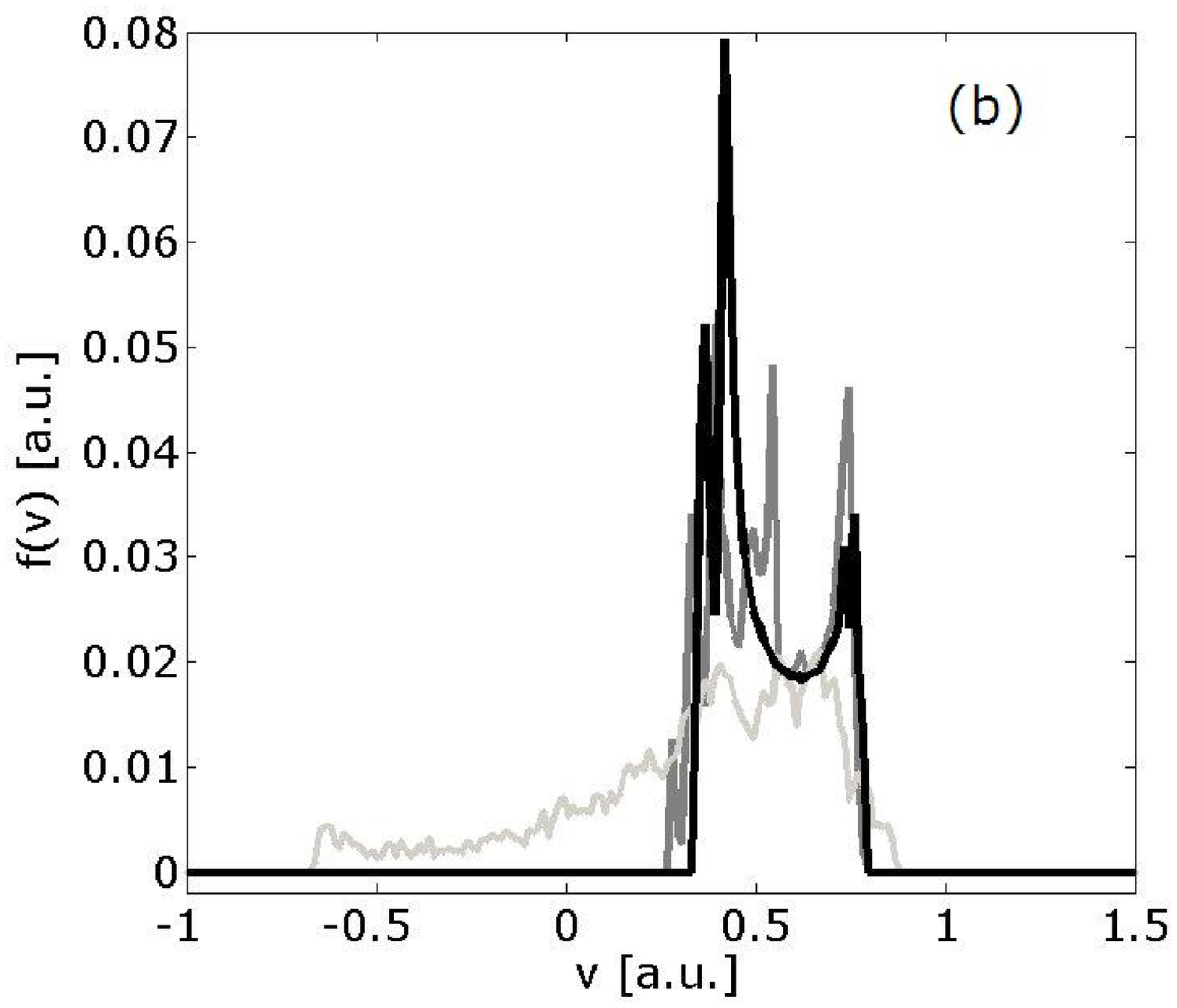}
\caption{(a): initial beam velocity distribution function for
$s=0.85$. (b): final beam velocity distribution function without
control term~(\ref{eqn:fTWT}) (light gray line), with the exact
control term (black line) and with the approximate control
term~(\ref{eqn:fsimappTWT})(dark gray line).} \label{Distr_sp85}
\end{figure}

\begin{figure}
\includegraphics[width=8.0cm, height=7.0cm]{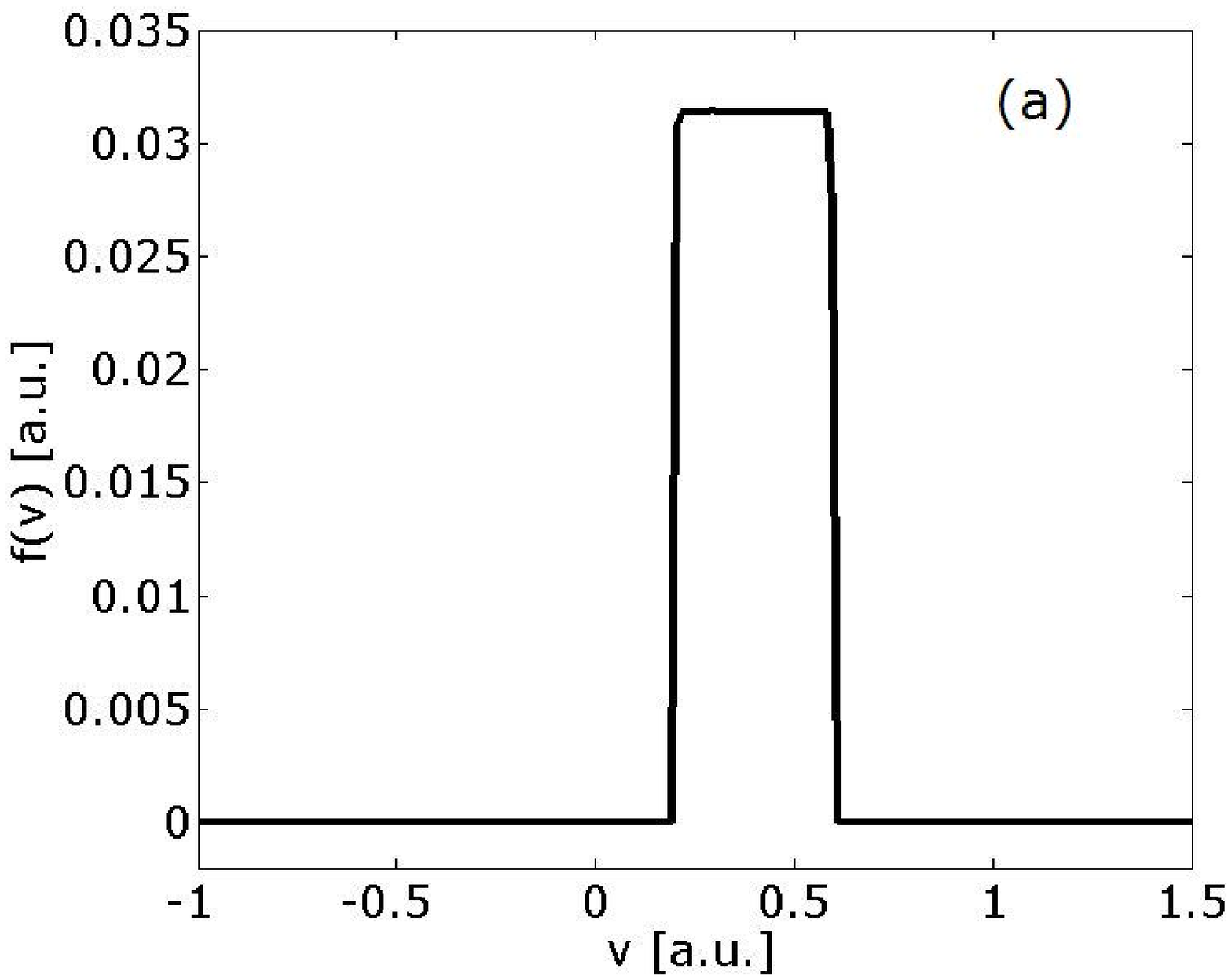}
\includegraphics[width=8.0cm, height=7.0cm]{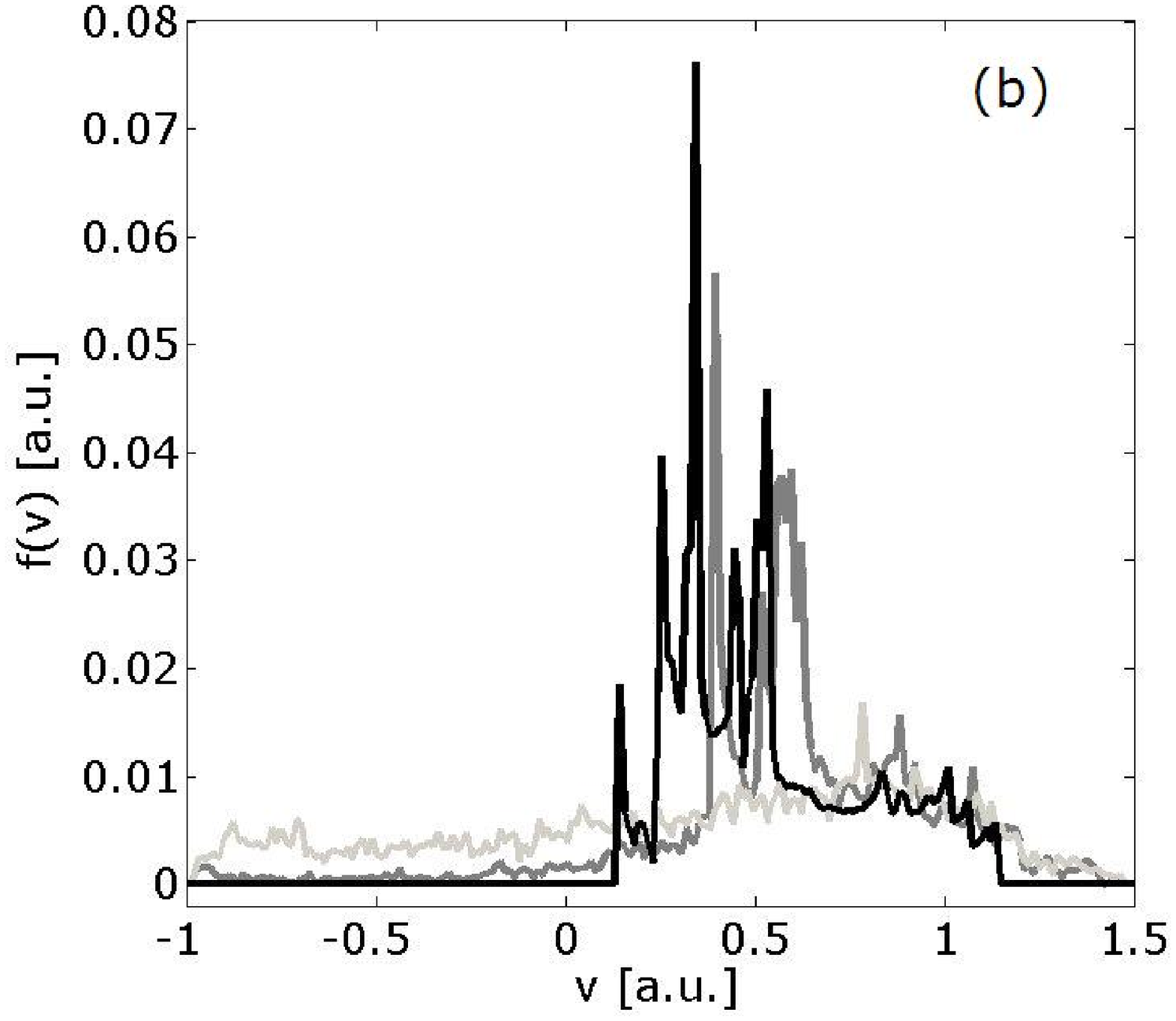}
\caption{(a): initial beam velocity distribution function for
$s=1.27$. (b): final beam velocity distribution function without
control term~(\ref{eqn:fTWT}) (light gray line), with the exact
control term (black line) and with the approximate control
term~(\ref{eqn:fsimappTWT}) (dark gray line).} \label{Distr_s1p27}
\end{figure}

In Figs.~\ref{Distr_sp85} and ~\ref{Distr_s1p27}, the initial
velocity distribution function of a set of $10^4$ particles is
compared with the final one obtained by integrating over a time
$t=50$, the dynamics governed by Hamiltonian (\ref{Ham:sim})
without control terms, plus the exact control
term~(\ref{eqn:fsimTWT}) and plus the approximate control
term~(\ref{eqn:fsimappTWT}). This investigation is performed for
the two different values of the Chirikov parameter, $s=0.85$ for
Fig.~\ref{Distr_sp85} and $s=1.27$ for Fig.~\ref{Distr_s1p27}. In
the case without any control term the original kinetic coherence
of the beam is lost which means that some electrons can have
velocities in the whole range in between resonances, $v\in [-0.67,
0.88]$. Adding the exact control term the particles are confined
in a selected region of phase space by the reconstructed invariant
tori, and the beam recovers a large part of its initial kinetic
coherence. For $s=0.85$, velocities of the electrons are now
between $0.33$ and $0.79$. The exact control term is also
efficient in the fully chaotic regime ($s=1.27$). Concerning the
approximate control term it is very efficient for $s=0.85$ while
its efficiency is smaller in the strongly chaotic regime. However,
it has still some regularizing effect, inducing the reconstruction
of stable islands in phase space which can catch and thus confine
a portion of the initial beam particles.

\subsection{Robustness of the method}
\label{loc:robust} The robustness of the control method for the
case $s=1.27$ is studied with respect to an error on the phase or
on the amplitude of the computed control term. In experiment,
given the frequency $\omega_1+\omega_2$, the wave number $k_1+k_2$
of the control term does not satisfy in general the dispersion
relation $k=K(\omega)$ since the dispersion relation is not
linear. In our case it means that the experimentally implemented
control term is not the exact one. For this reason we investigate
the robustness of the control term given by
Eq.~(\ref{eqn:fsimappTWT}) with a phase error $\varphi$, that is
\begin{equation}
  f(x,t)= 2\varepsilon^2r\cos[(k+1)x-kt+\varphi],
    \label{eqn:fappfasi}
\end{equation}
and with an error on its amplitude ruled by a factor $\delta$,
that is
\begin{equation}
  f(x,t)= 2\varepsilon^2r\delta\cos[(k+1)x-kt].
    \label{eqn:fappAmp}
\end{equation}
The values given by Eq.~(\ref{eqn:fsimappTWT}) are $\varphi=0$ and
$\delta=1$. In order to quantify the robustness of the approximate
control term given by Eq.~(\ref{eqn:fappfasi}) or
Eq.~(\ref{eqn:fappAmp}), we introduce the kinetic coherence
indicator defined as the ratio of the variance of the initial beam
over the variance of the distribution function after a given
integration time. The number of particles, the integration time
and the initial conditions are equal to the ones used in the
previous section.

\begin{figure}
\includegraphics[width=7.9cm,height=7.0cm]{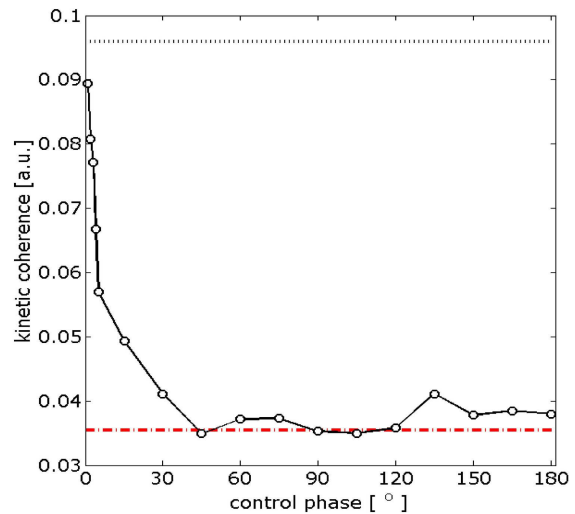}
\caption{Kinetic coherence versus the phase introduced in the
approximate control term given by Eq.~(\ref{eqn:fappfasi}). The
red dash-dotted line indicates the kinetic coherence value for the
non-controlled case and the dotted line the kinetic coherence
value for the approximate control term. The overlap parameter is
$s=1.27$.} \label{fig:kin_fase}
\end{figure}

In Fig.~\ref{fig:kin_fase} we show the kinetic coherence as a function of
the phase of the approximate control term for the strongly chaotic regime $s=1.27$. We
notice that $\varphi$ or $-\varphi$ will give the same velocity distribution function for
symmetry reason. Therefore we only consider the range $\varphi\in[0,\pi]$. The
efficiency of the
approximated control term is very sensitive with respect to the phase. In
fact an error of $5^{\mathrm o}-6^{\mathrm o}$ causes a decrease of the kinetic coherence of about $50\%$
and with an error greater than $30^{\mathrm o}$ the kinetic coherence drops in the range of
values of the non-controlled case.

\begin{figure}
\includegraphics[width=8.3cm,height=7.2cm]{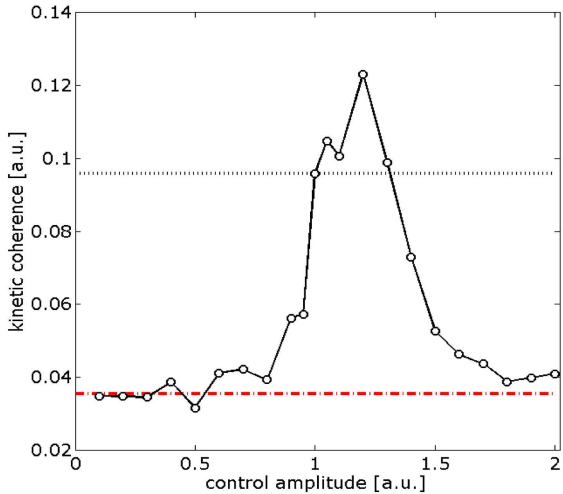}
\caption{Kinetic coherence versus $\delta$ for the approximate
control term given by Eq.~(\ref{eqn:fappAmp}). The red dash-dotted
line indicates the kinetic coherence value for the non-controlled
case and the dotted line the kinetic coherence value for the
approximate control term given by Eq.~(\ref{eqn:fsimappTWT}).The
overlap parameter is $s=1.27$.} \label{fig:kin_amp}
\end{figure}

Concerning the robustness with respect to an error on the
amplitude of the approximate control term, we plot on
Fig.~\ref{fig:kin_amp} the behavior of the kinetic coherence as a
function of the $\delta$-factor which multiplies the amplitude of
the approximate control term. We notice that around the reference
value of $\delta=1$ (no error) there is a region
($\delta\in[1,1.3]$) where the approximate control term is very
efficient in confining the beam of test particles with a kinetic
coherence in between $[0.096,0.12]$. On the other hand reducing
the amplitude of the control term, i.e. its energy, there is a
region where one has still a confining effect on the beam
particle. For example for $\delta=0.6$ the kinetic coherence is
larger by $50\%$ of the value of the non-controlled case.

\section{Experimental tests}
\label{experiment}
\subsection{Experimental set-up}
\label{experiment:setup}
\begin{figure}[tbp]
\includegraphics[width=7.5cm,height=4.9cm]{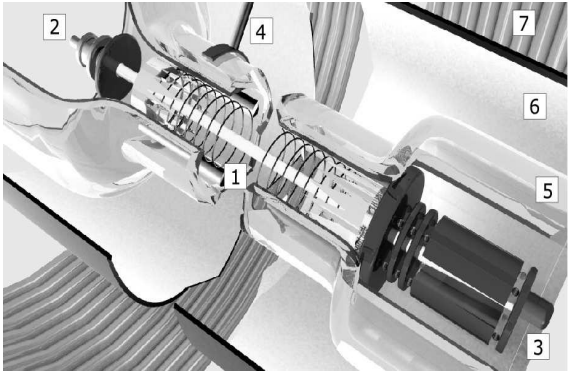}
\caption{Sketch of the Travelling Wave Tube: (1) helix, (2)
electron gun, (3) trochoidal analyzer, (4) antenna, (5) glass
vacuum tube, (6) slotted rf ground cylinder, and (7) magnetic
 coil.}
\label{fig:twt}
 \end{figure}

\begin{figure}[tbp]
\includegraphics[width=7.5cm,height=7.2cm]{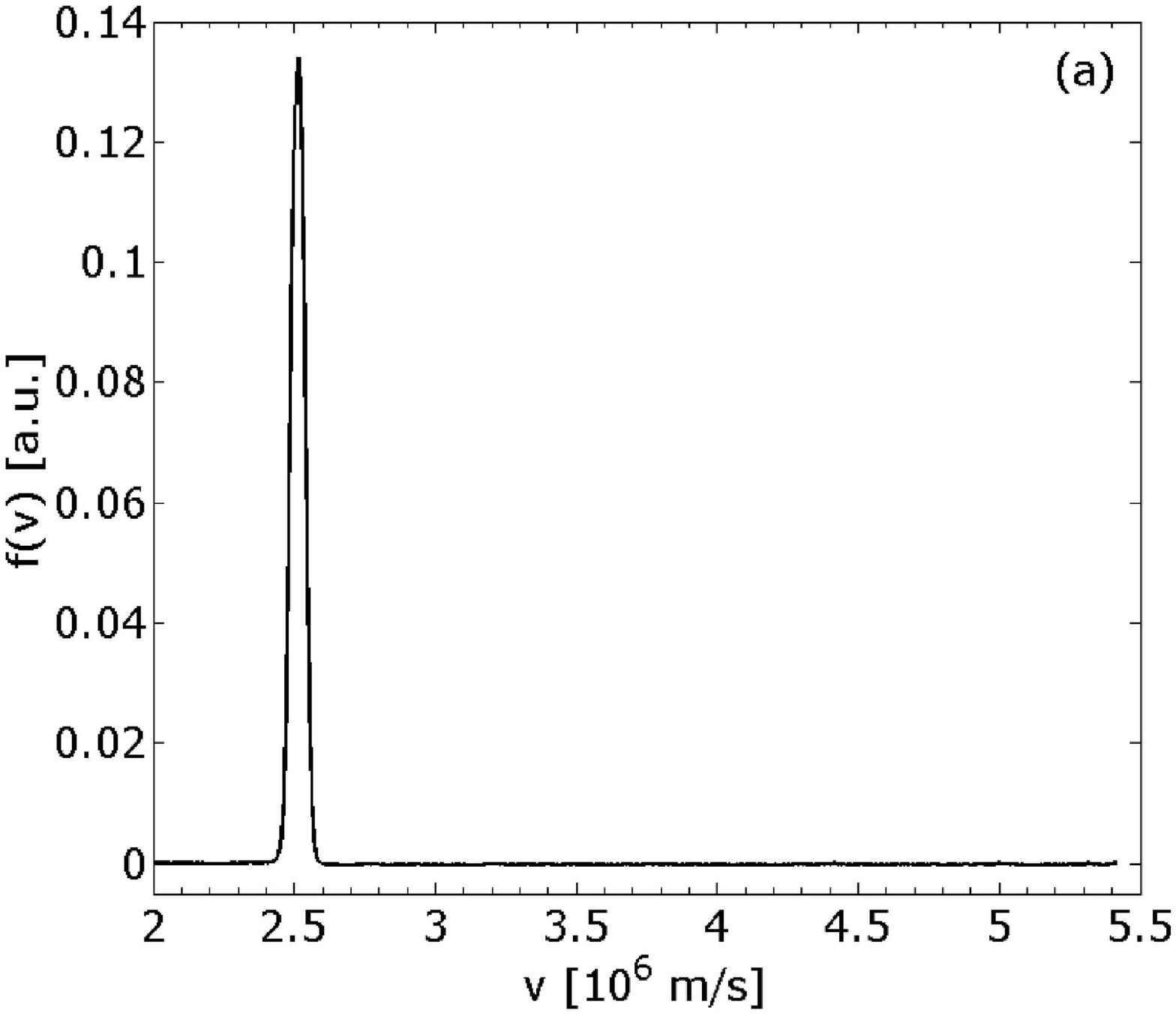}
\includegraphics[width=7.5cm,height=7.2cm]{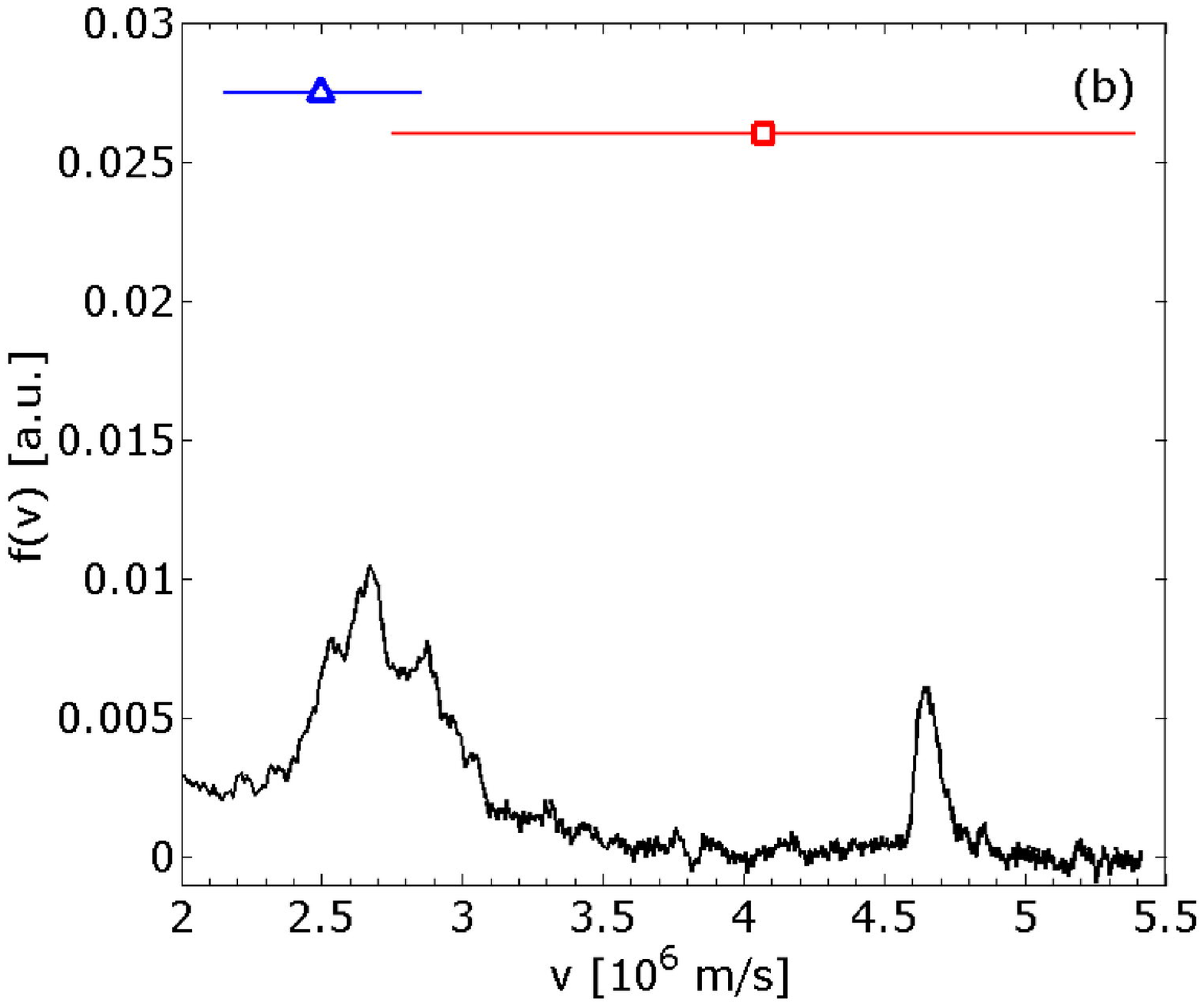}
\includegraphics[width=7.5cm,height=7.2cm]{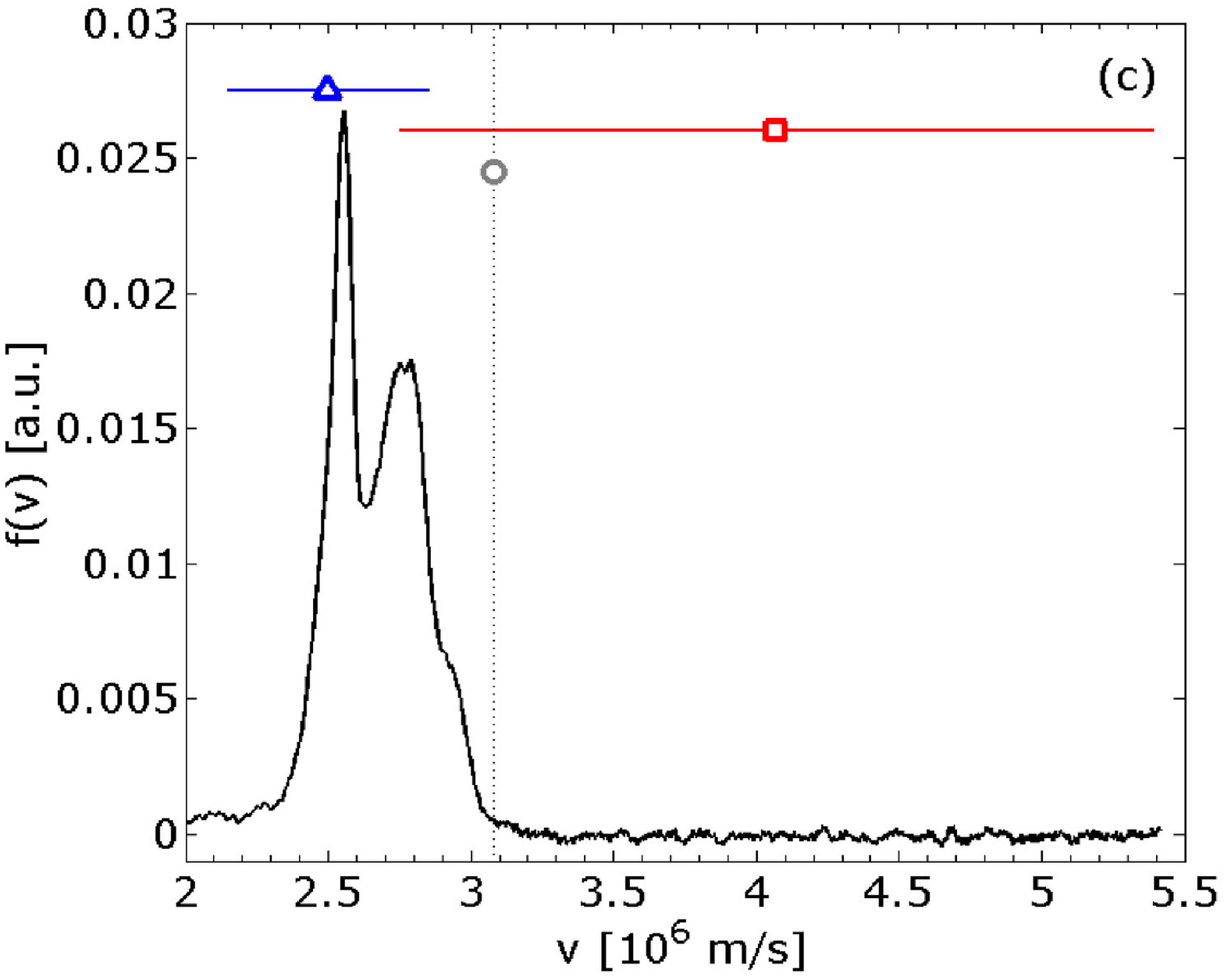}
\caption{Beam velocity distribution functions at the output of the
TWT for $s = 0.85$ : (a) test beam ($I_b=50$ nA) without
electrostatic wave, (b) with helix mode (trapping domain shown by
red line with phase velocity marked by a square) and beam mode
(trapping domain shown by blue line with phase velocity marked by
a triangle) at 30 MHz , (c) with an additional controlling wave at
$60$ MHz and phase velocity given by grey circle and dotted
line.}\label{fig:085}
\end{figure}

\begin{figure}[tbp]
\includegraphics[width=7.5cm,height=7.2cm]{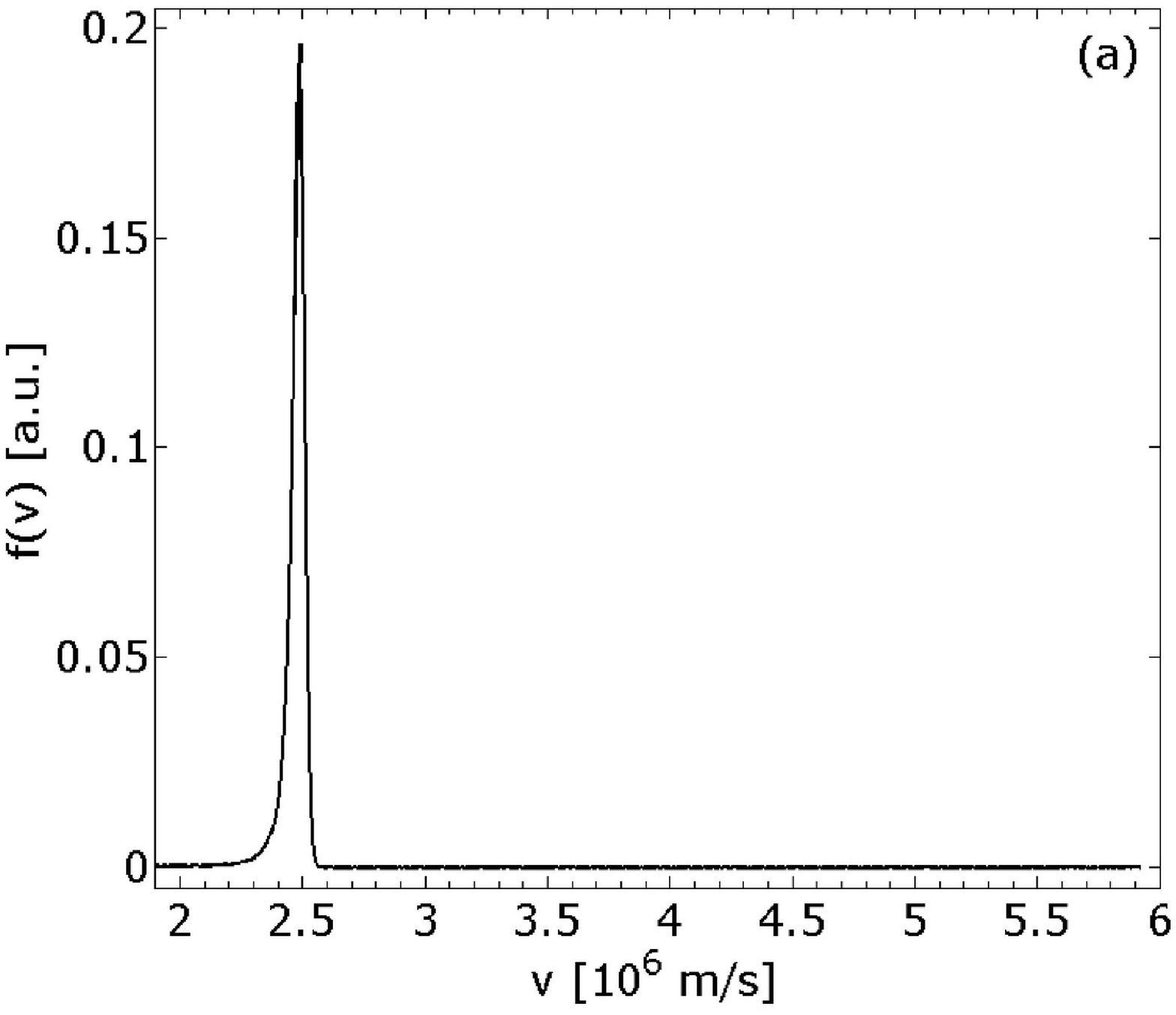}
\includegraphics[width=7.5cm,height=7.2cm]{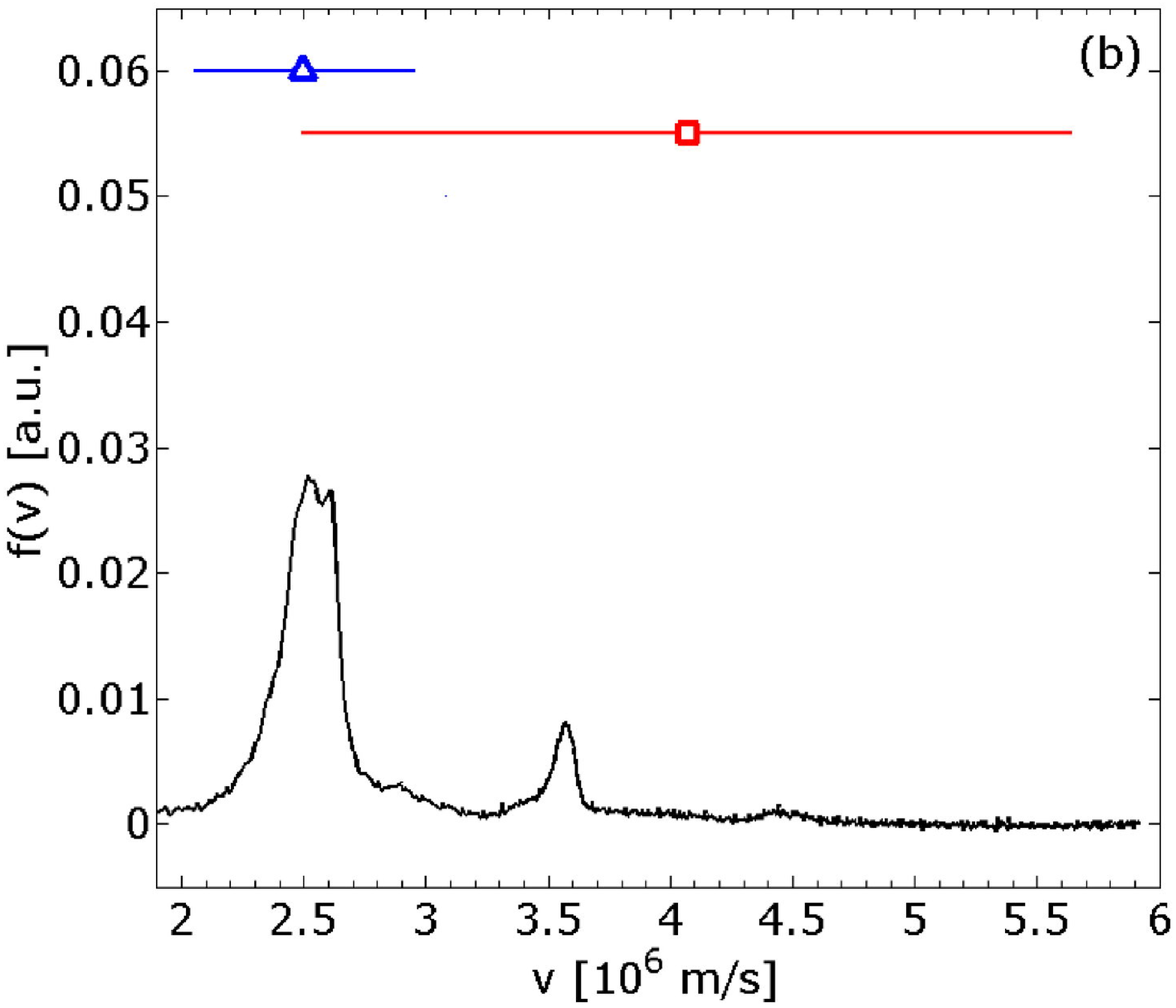}
\includegraphics[width=7.5cm,height=7.2cm]{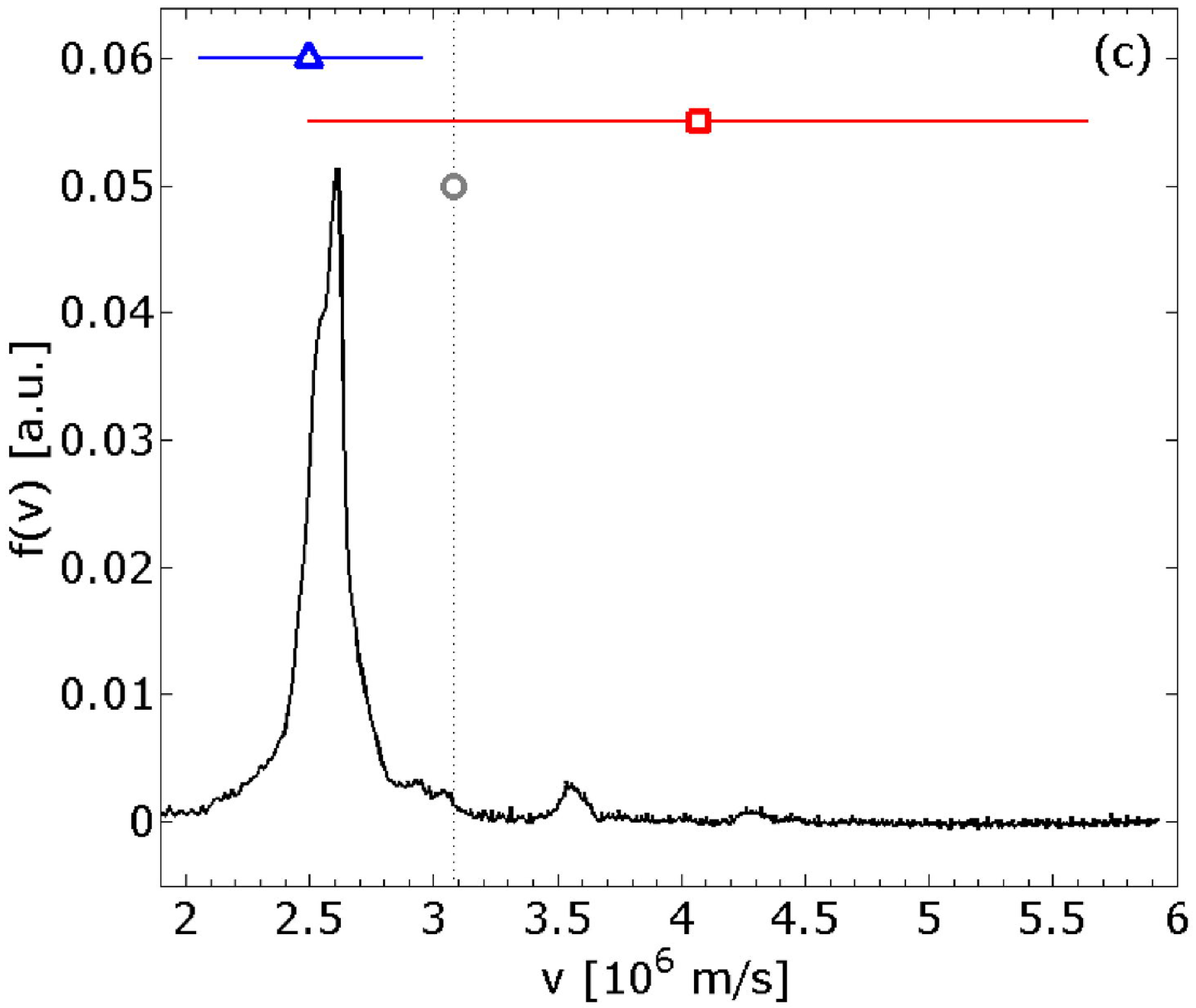}
\caption{Beam velocity distribution functions at the output of the
TWT for $s = 1.27$ : (a) test beam ($I_b=2$ nA) without
electrostatic wave, (b) with helix mode (trapping domain shown by
red line with phase velocity marked by a square) and beam mode
(trapping domain shown by blue line with phase velocity marked by
a triangle) at 30 MHz, (c) with an additional controlling wave at
$60$ MHz and phase velocity given by grey circle and dotted
line.}\label{fig:127}
\end{figure}

 The experimental implementation of the control term is performed in a
long traveling wave tube (TWT)~\cite{Pierce,Gilmour} extensively
used to mimic beam plasma interaction~\cite{Tsunoda,Dimonte} and
recently to observe resonance overlap responsible for Hamiltonian
chaos ~\cite{Macor1}. The TWT sketched in Fig.~\ref{fig:twt} is
made up of three main elements: an electron gun, a slow wave
structure (SWS) formed by a helix with axially movable antennas,
and an electron velocity analyzer. The electron gun creates a beam
which propagates along the axis of the SWS and is confined by a
strong axial magnetic field with a typical amplitude of $0.05
{\,\rm T}$ which does not affect the axial motion of the
electrons. The central part of the gun consists of the
grid-cathode subassembly of a ceramic microwave triode and the
anode is replaced by a Cu plate with an on-axis hole whose
aperture defines the beam diameter equal to $1 {\,\rm mm}$. Beam
currents, $I_b < 1 {\,\rm mA}$, and maximal cathode voltages,
$|V_c| < 200 {\,\rm V}$, can be set independently; an example of
typical velocity distribution functions is given in
Figs.~\ref{fig:085} and \ref{fig:127} (panel (a)). Two correction
coils provide perpendicular magnetic fields to control the tilt of
the electron beam with respect to the axis of the helix. For the
data shown in this article $I_b$ is chosen weak enough to ensure
that the beam induces no wave growth and the beam electrons can be
considered as test electrons. The SWS is long enough to allow
nonlinear processes to develop. It consists in a wire helix that
is rigidly held together by three threaded alumina rods and is
enclosed by a glass vacuum tube. The pressure at the ion pumps on
both ends of the device is $2 \times10^{-9} {\,\rm Torr}$. The $4
{\,\rm meter}$ long helix is made of a $0.3 {\,\rm mm}$ diameter
Be-Cu wire; its radius is equal to $11.3 {\,\rm mm}$ and its pitch
to $0.8 {\,\rm mm}$. A resistive rf termination at each end of the
helix reduces reflections. The maximal voltage standing wave ratio
is 1.2 due to residual end reflections and irregularities of the
helix. The glass vacuum jacket is enclosed by an axially slotted
$57.5 {\,\rm mm}$ radius cylinder that defines the rf ground.
Inside this cylinder but outside the vacuum jacket are four
axially movable antennas which are capacitively coupled to the
helix and can excite or detect helix modes in the frequency range
from 5 to 95 {\,\rm MHz}. Only the helix modes are launched, since
empty waveguide modes can only propagate above $2 {\,\rm GHz}$.
These modes have electric field components along the helix axis
~\cite{Dimonte}. Launched electromagnetic waves travel along the
helix with the speed of light; their phase velocities, ${v_{\phi
j}}$, along the axis of the helix are smaller by approximately the
tangent of the pitch angle, giving $2.8 \times10^6 {\,\rm m/s} <
{v_{\phi j}} < 5.3 \times10^6 {\,\rm m/s}$. Waves on the beamless
helix are slightly damped, with $|{k_{j}^{0i}}|/|{k_{j}^{0r}}|
\approx 0.005$ where ${k^0 = k^{0r} + {\rm i\,} k^{0i}}$ is the
beamless complex wave number. The dispersion relation closely
resembles that of a finite radius, finite temperature plasma, but,
unlike a plasma, the helix does not introduce appreciable noise.
Finally the cumulative changes of the electron beam distribution
are measured with the velocity analyzer, located at the end of the
interaction region. This trochoidal analyzer~\cite{guyomarc'h}
works on the principle that electrons undergo an ${\bf E \times
B}$ drift when passing through a region in which an electric field
${\bf E}$ is perpendicular to a magnetic field ${\bf B}$. A small
fraction (0.5\%) of the electrons passes through a hole in the
center of the front collector, and is slowed down by three
retarding electrodes. Then the electrons having the correct drift
energy determined by the potential difference on two parallel
deflector plates are collected after passing through an off-axis
hole at the back of the analyzer. The time averaged collected
current is measured by means of a pico-ampermeter. Retarding
potential and measured current are computer controlled, allowing
an easy acquisition and treatment with an energy resolution lower
than $0.5 {\,\rm eV}$. In the absence of any emitted wave, after
propagating along the helix, the beam exhibits a sharp velocity
distribution function with a velocity width mainly limited by the
analyzer resolution as shown in Figs.~\ref{fig:085}
and~\ref{fig:127} (panel (a)). For Fig.~\ref{fig:085}a, the beam
with radius 3mm is diffracted by passing through the three
grounded grids of a spreader~\cite{spreader} just after leaving
the gun while for Fig.~\ref{fig:127}a, the beam radius is $1
{\,\rm mm}$ and the spreader has beem removed for the sake of
simplicity.

\begin{figure}[tbp]
\includegraphics[width=7.5cm]{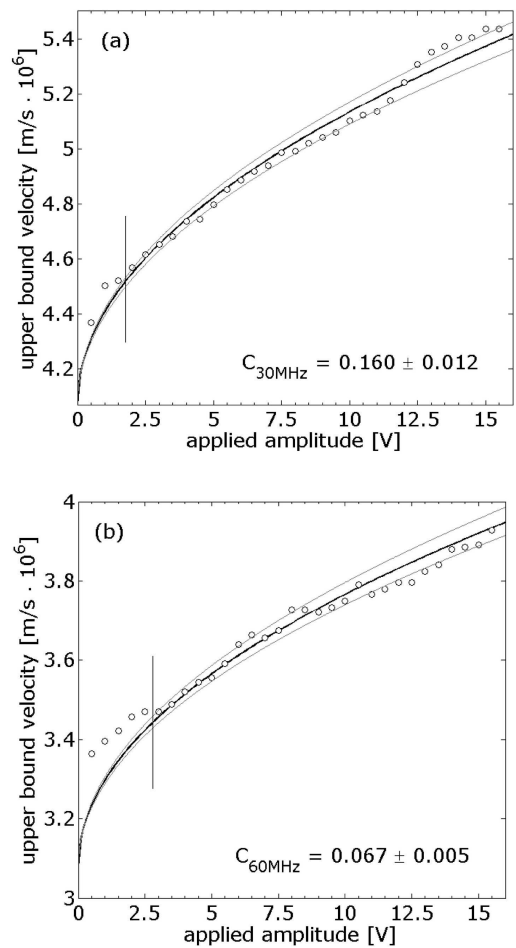}
\caption{Measure of helix mode coupling constant $C_h$ calculated
from the measured upper bound velocity of trapping domain
(circles) at frequency $ f $ equal to (a) 30{\,\rm MHz}  and (b)
60{\,\rm MHz}. Black line parabola is obtained using the average
value of $ C_h $. Gray parabolas are given by the average error.
The parabolic fit is only valid beyond the vertical segment
indicating the wave amplitude beyond which beam trapping occurs
over the length of the TWT.} \label{fig:const3060}
\end{figure}

\begin{figure}[tbp]
\includegraphics[width=7.5cm,height=7.2cm]{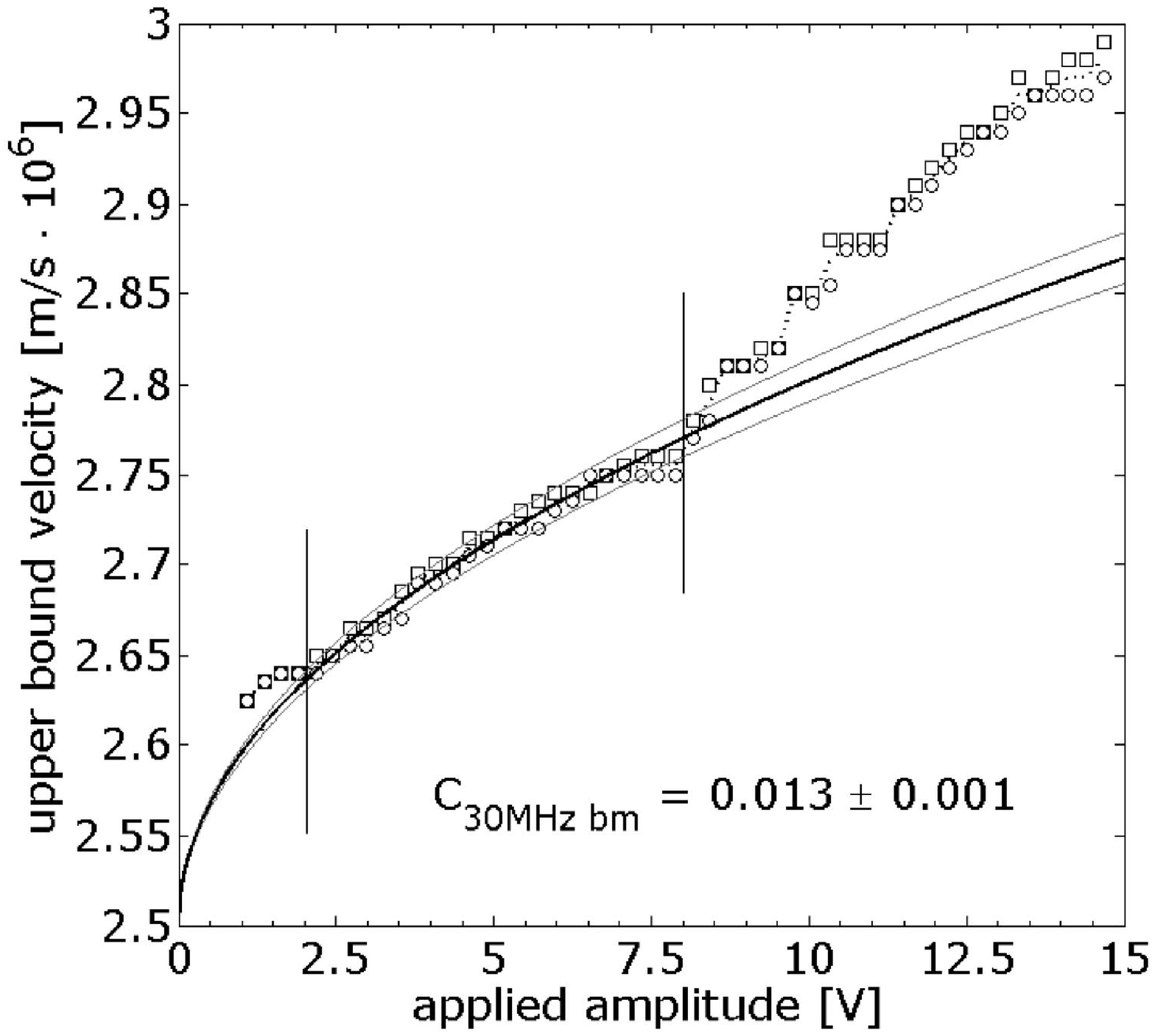}
\caption{Measure of coupling constant $C_b$, at frequency equal to
30{\,\rm MHz} for a beam mode at $2.5\times 10^6${\,\rm m/s}.
Black line parabola is obtained using the average value of $ C_b
$, calculated from the upper bound velocity of the trapping domain
between the vertical segments where the beam become trapped over
the length of the TWT and the "devil's staircase" is not yet
evident~\cite{escalier}. Two independent measures are shown by
circles and squares to give an error estimate.}
\label{fig:const30b}.
\end{figure}

\subsection{Experimental implementation of the control term}
\label{experiment:implementation} We apply an oscillating signal
at the frequency of 30 MHz on one antenna. It generates two waves:
a helix mode with a phase velocity equal to
$v_{\varphi}=4.07\times 10^6$ m/s, a beam mode with a phase
velocity equal to the beam velocity $v_b$ (in fact two modes with
pulsation $\omega=kv_b\pm\omega_b$ corresponding to the beam
plasma mode with pulsation $\omega_b=(n_be^2/m\epsilon_0)^{1/2}$,
Doppler shifted by the beam velocity $v_b$, merging in a single
mode since $\omega_b\ll\omega$ in our conditions).
Figures~\ref{fig:085} and~\ref{fig:127} (panel (b)) show the
measured velocity distributions of the beam after interacting with
these two modes over the length of the TWT for two different
values of the Chirikov parameter. The case with $s=0.85$ was
previously investigated ~\cite{c18}. The red square (resp. blue
triangle) shows the phase velocity $v_{\varphi}$ (resp.$v_{b}$) of
the helix (resp. beam) mode on the middle of the resonant domain
determined as the trapping velocity width of the helix  mode
$v_{\varphi}\pm 2\sqrt{eC_h\Phi/m}$ (resp.~$v_b\pm
2\sqrt{eC_{b}\Phi/m}$~) where $\Phi$ is the signal amplitude
applied on the antenna and $C_h\Phi=3542 {\rm mV}$ (resp.
$C_{b}\Phi=286 {\rm mV}$) is the real amplitude of the helix
(resp. beam) mode. Both $C_h$ and $C_b$ are determined
experimentally by the estimations of the coupling constant  for
helix $C_h$(resp. beam $C_b$) mode. As shown in
Fig.~\ref{fig:const3060} the helix mode coupling coefficient $C_h$
is obtained by fitting a parabola through the measured upper bound
velocity (circles) after the cold test beam with initial velocity
equal to the wave phase velocity has been trapped by the wave at a
given frequency over the total length of the TWT. As shown in
Fig.~\ref{fig:const30b}, the beam mode coupling coefficient $C_b$
is obtained by fitting a parabola through the measured upper bound
velocity (circles) for a beam with a mean velocity very different
from the helix mode phase velocity at the considered frequency.
These two domains overlap and the break up of invariant KAM tori
(or barriers to velocity diffusion) results in a large spread of
the initially narrow beam of Figs.~\ref{fig:085} and~\ref{fig:127}
(panel (b)) over the chaotic region ~\cite{Macor1}. We now use an
arbitrary waveform generator~\cite{Tsunoda} to launch the same
signal at $30$ MHz and an additional control wave with frequency
equal to $60$ MHz, an amplitude and a phase given by
Eq.~(\ref{eqn:f2app}). The beam velocity is also chosen in such a
way that the wave number of the helix mode at 60 MHz properly
satisfies the dispersion relation function shown as circles in
Fig.~\ref{fig:reldisp}. We neglect the influence of the beam mode
at $60$ MHz since its amplitude is at least an order of magnitude
smaller than the control amplitude as shown by comparing Figs.
~\ref{fig:const3060}a and ~\ref{fig:const30b} for $30$ MHz. As
observed on Figs.~\ref{fig:085} and~\ref{fig:127}(panel (c)) where
the grey circle indicates the phase velocity of the controlling
wave, the beam recovers a large part of its initial kinetic
coherence. For $s=0.85$ (see Fig.~\ref{fig:085}c) the beam does
not spread in velocity beyond the reconstructed KAM tori, in
agreement with the numerical simulations of Fig.~\ref{fig:Poinc1}.
For the more chaotic regime (see Fig.~\ref{fig:127}b) with
$s=1.27$ the improvement of the kinetic coherence is still present
as shown in Fig.~\ref{fig:127}c. It can no more be associated with
the reconstruction of a local velocity barrier, as expected from
the numerical results in Fig.~\ref{fig:Poinc2} (panel (c)).  For
this last overlap parameter an experimental exploration of the
robustness of the method will be shown in the next section.

\subsection{Robustness of the method}
\label{experiment:robust} In our experiment the control term is
given by an additional wave whose frequency, amplitude and phase
are  computed as shown in Sec.~\ref{loc:control}. In order to
quantify the robustness of the method we will compare the various
experimental situations to a reference one (Fig.~\ref{fig:127}a).
 This reference is taken as the (initial) cold
beam distribution function. An example of the distribution we were
able to reach with control is given in Fig.~\ref{fig:127}c. The
control amplitude is $140${\,\rm mV} in agreement with $144${\,\rm
mV} given by the method up to experimental errors. The phase is
chosen experimentally and arbitrarily labelled $0^{\rm o}$. The
beam velocity is chosen equal to $2.498\times 10^6${\,\rm m/s} in
agreement with $2.51\times 10^6${\,\rm m/s} as estimated from the
dispersion relation shown in Fig.~\ref{fig:reldisp}.

We investigate the robustness of the control method with respect
to variation of phase and amplitude in the approximate control
term given by Eq.~(\ref{eqn:f2app}).

\begin{figure}[tbp]
\includegraphics[width=7.5cm,height=7.2cm]{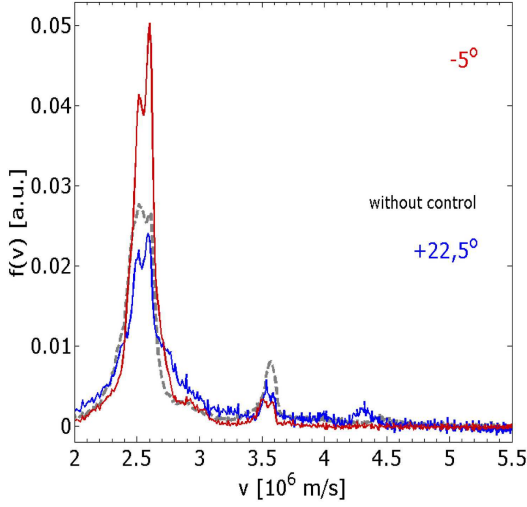}
\caption{Distribution functions for the maximum kinetic coherence
value (resp. minimum), blue line (resp. red line), is compared
with the non-controlled distribution function (grey dashed line).
The overlap parameter is $s=1.27$.} \label{fig:varfaseb}
\end{figure}

\begin{figure}[tbp]
\includegraphics[width=7.5cm,height=7.2cm]{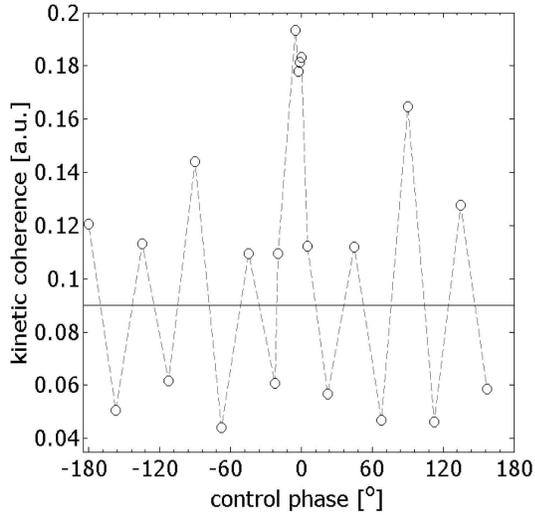}
\caption{Kinetic coherence defined as the variance of the cold
beam distribution function  over the variance of distribution
function for different control term phases, an optimized amplitude
$140${\,\rm mV} and a beam velocity $2.498\times 10^6${\,\rm m/s}.
Control becomes efficient in a narrow region close to zero. Solid
line shows the level of the kinetic coherence for the
non-controlled case. Dashed line is only an eye-guide. The overlap
parameter is $s=1.27$.}\label{varsupfase}
\end{figure}

We use the kinetic coherence indicator to quantify the effect of
the control, defined as the ratio of variance of the cold beam
distribution function over the variance of the distribution
function. Other indicators (integral and uniform distances) were
used and gave similar results. Figure~\ref{fig:varfaseb} shows the
velocity distribution functions for two values of the phase
($-5^{\rm o}$ and $22.5^{\rm o}$) keeping the other parameters
constant. It shows that for a phase equal to $-5^{\rm o}$ close to
the reference value the two velocity distribution functions are
very similar, and more peaked at $-5^{\rm o}$ than at $0^{\rm o}$.
For $22.5^{\rm o}$, the control wave has the opposite effect,
increasing chaos. In Fig.~\ref{varsupfase} we show the kinetic
coherence as a function of the phase of the approximate control
term. It shows a narrow region around the reference value where
the control wave is the most efficient.

\begin{figure}[tbp]
\includegraphics[width=7.5cm,height=7.2cm]{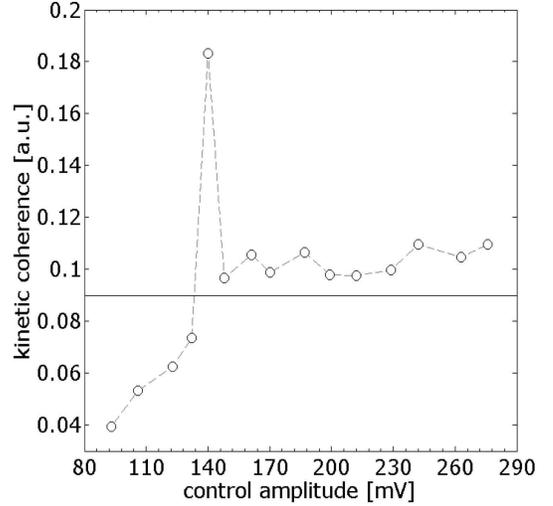}
\caption{Kinetic coherence defined such as in
Fig.~\ref{varsupfase}a for different control amplitudes and a
phase $0^{\rm o}$ and a beam velocity $2.498\times10^6$ {\,\rm
m/s}. Control becomes efficient in a narrow region around $140$
{\,\rm mV}. Solid line shows the level of the kinetic coherence
for the non-controlled chaos. Dashed line is only an eye-guide.
The overlap parameter is $s=1.27$.} \label{fig:var_int_ampli}
\end{figure}

\begin{figure}[tbp]
\includegraphics[width=7.5cm,height=7.2cm]{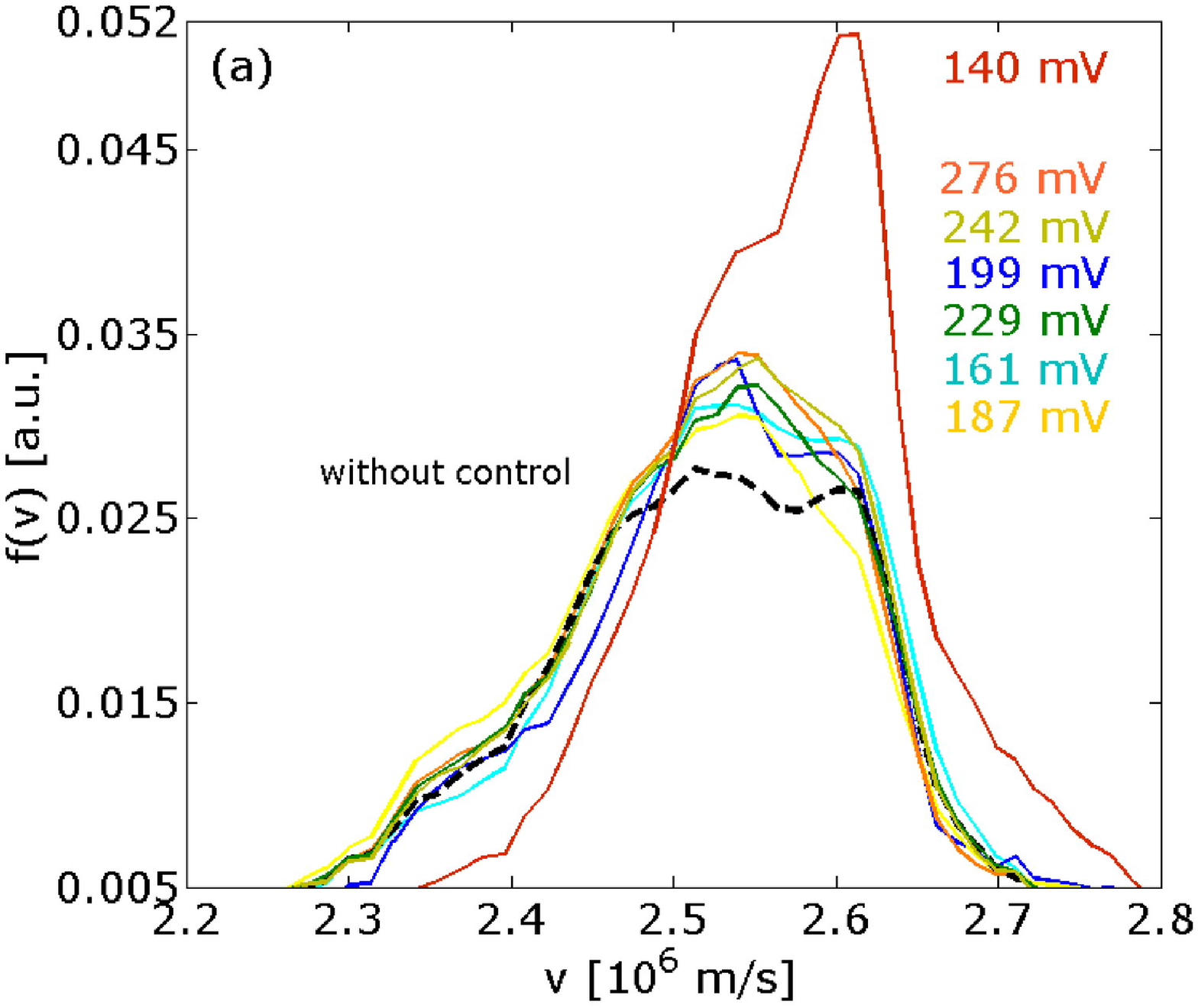}
\includegraphics[width=7.5cm,height=7.2cm]{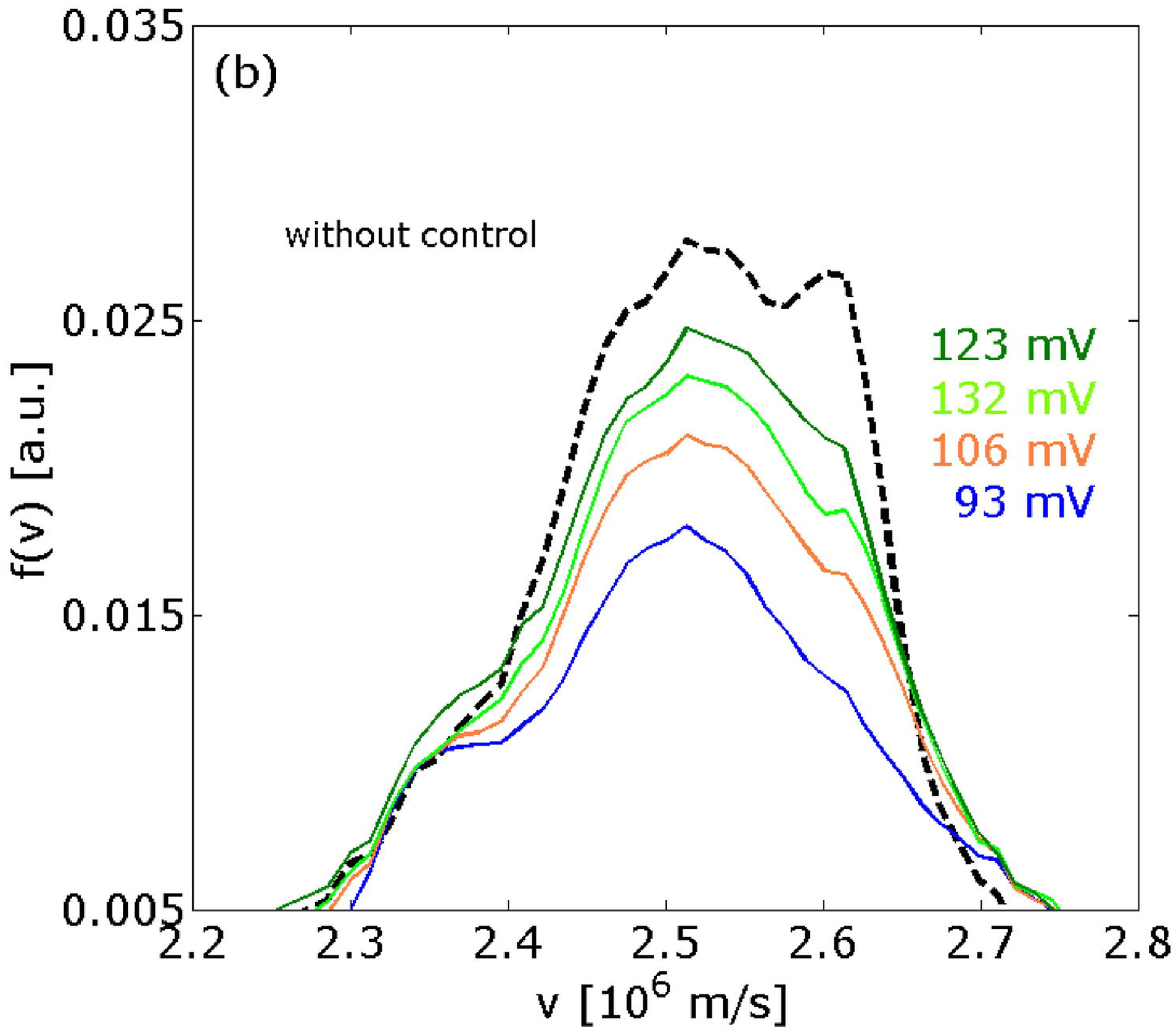}
\includegraphics[width=7.5cm,height=7.2cm]{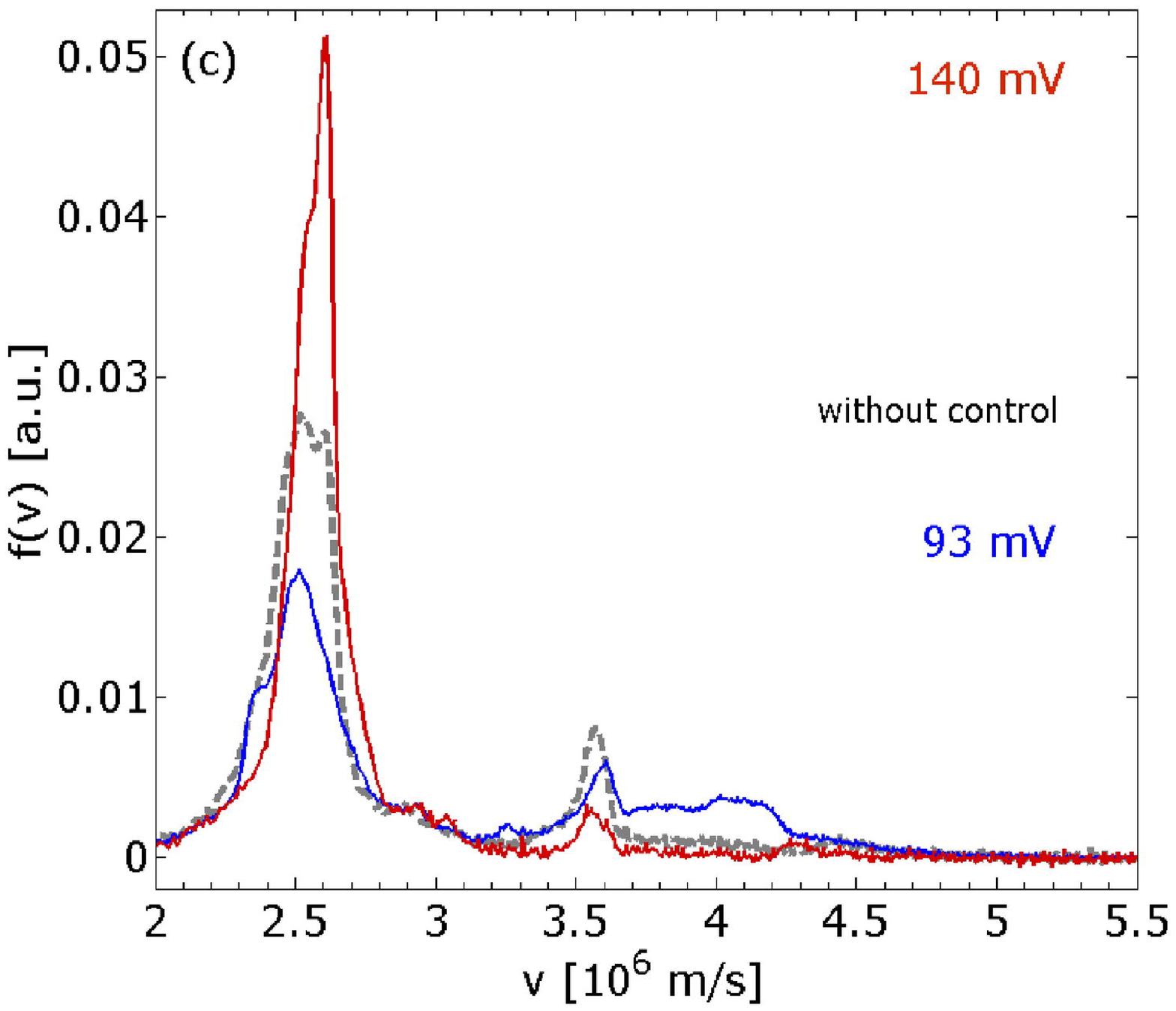}
\caption{Enlargement of velocity distribution functions close to
initial beam velocity for different control amplitudes: (a)
larger, (b) lower than experimentally optimized amplitude $140$
{\,\rm mV}. (c) velocity distribution functions for three
different control amplitudes at fixed control phase equal to
$0^{\rm{o}}$. The overlap parameter is $s=1.27$.}\label{fig:ampli-}
\end{figure}

In Fig.~\ref{fig:var_int_ampli}, we represent the kinetic
coherence as a function of the amplitude of the control wave. When
changing the control wave amplitude a resonance condition in a
narrow region around the optimized amplitude is still observed.
For amplitudes smaller than the reference (computed) value the
effect of the control decays fast and the electron velocities are
more widely spread than in the non-controlled case. Besides, for
larger values, the beam velocity spread increases but the control
term energy becomes comparable to the beam mode energy changing
radically the initial system. We have observed, due to beam
current conservation, a lower peak at initial beam velocity
implies that electron velocities are more widely spread. An
enlargement of distribution around the main peak is shown in
Figs.~\ref{fig:ampli-} a,b and confirms that $140${\,\rm mV}
appears to be the optimum.

\begin{figure}[tbp]
\includegraphics[width=7.5cm,height=7.2cm]{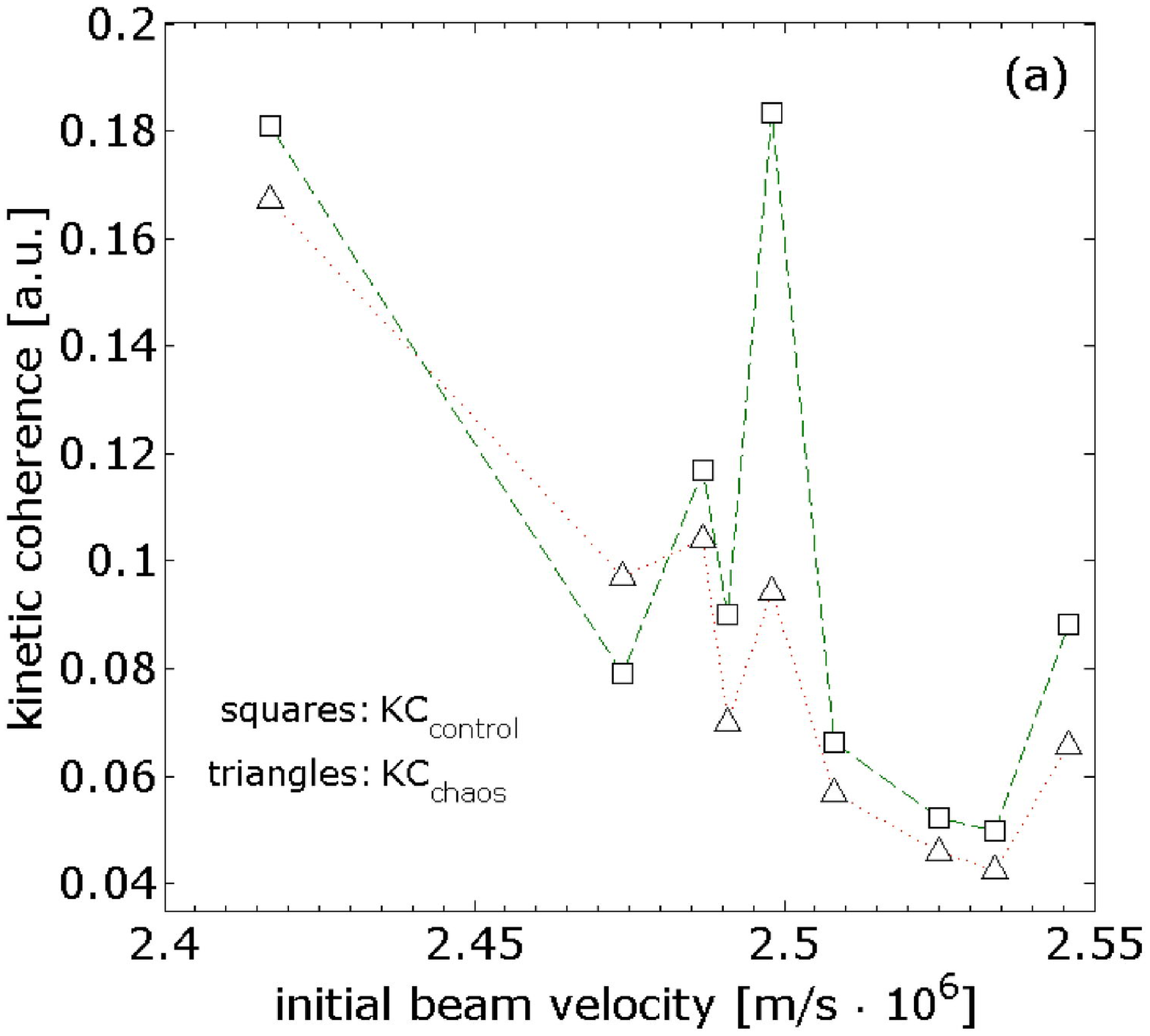}
\includegraphics[width=7.5cm,height=7.2cm]{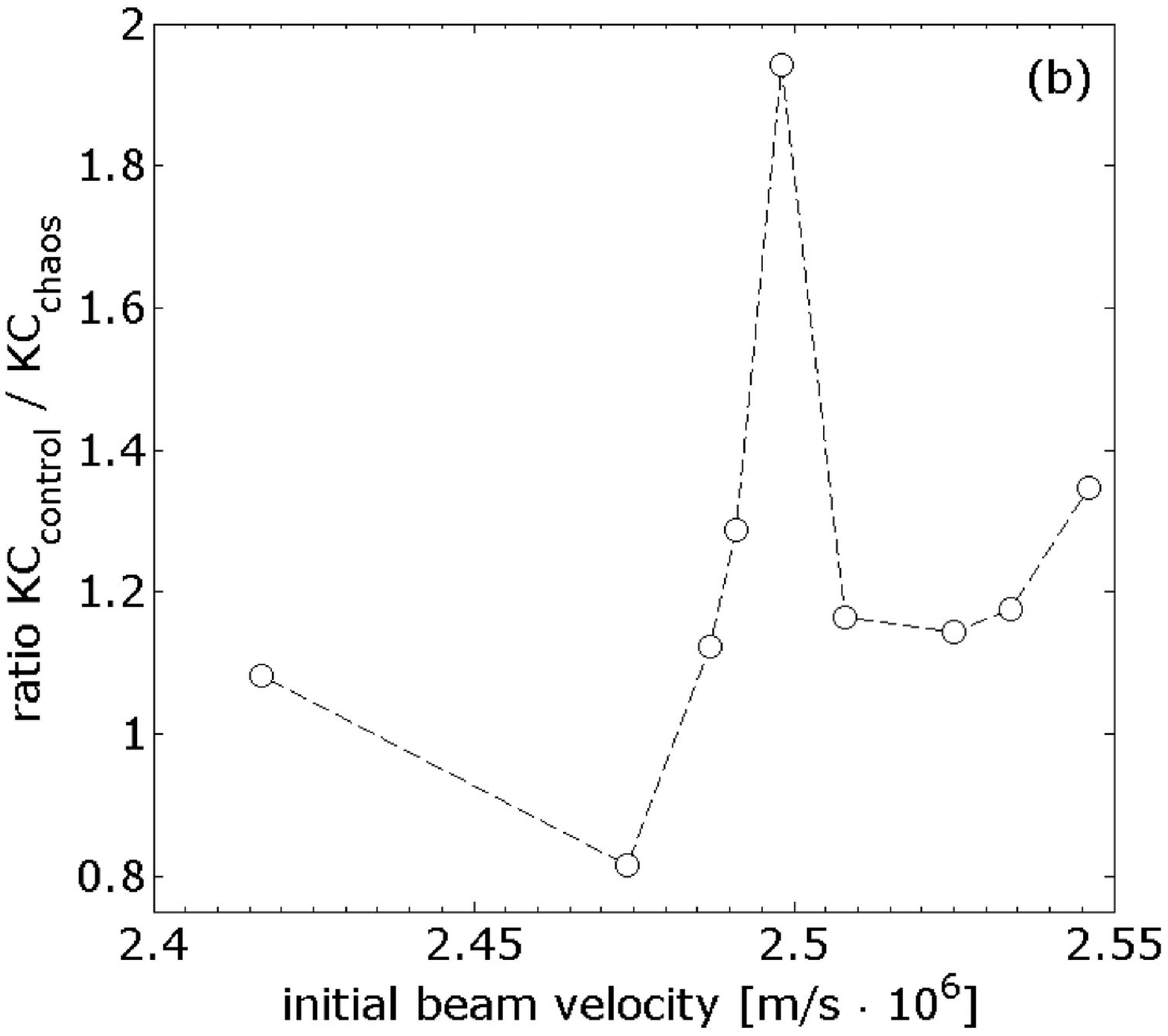}
\caption{Kinetic coherence versus initial beam velocity with
optimized control term amplitude ($140${\,\rm mV}) and  phase
($0^{\rm o}$). (a), squares (resp. triangles) show the values for
velocity distribution function obtained with (resp. without)
applied control term. (b), Ratio between kinetic coherence
measured with and without control for different initial beam
velocity. The overlap parameter is $s=1.27$.}\label{fig:varvelo}
\end{figure}

\begin{figure}[tbp]
\includegraphics[width=7.5cm,height=7.2cm]{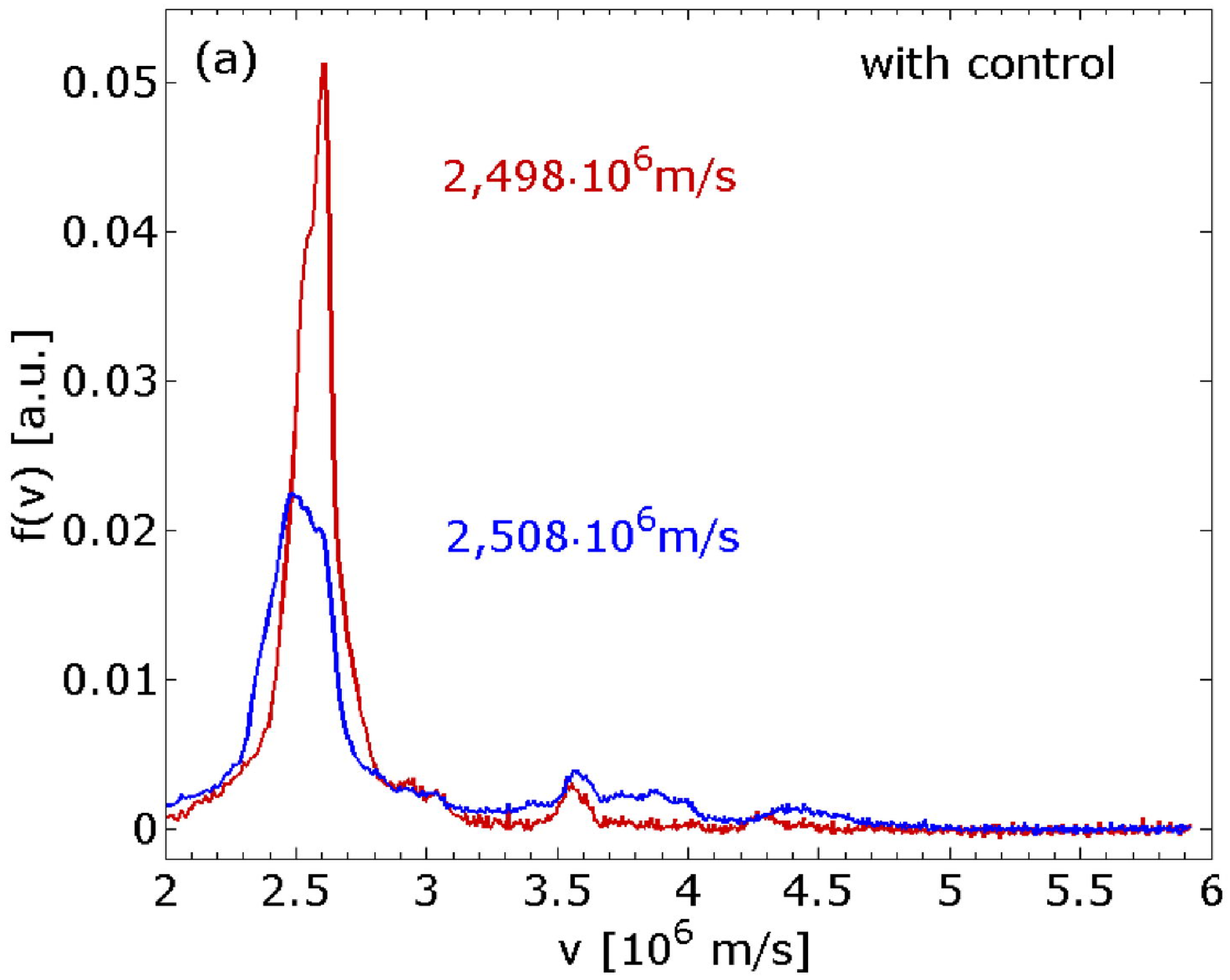}
\includegraphics[width=8cm,height=7.4cm]{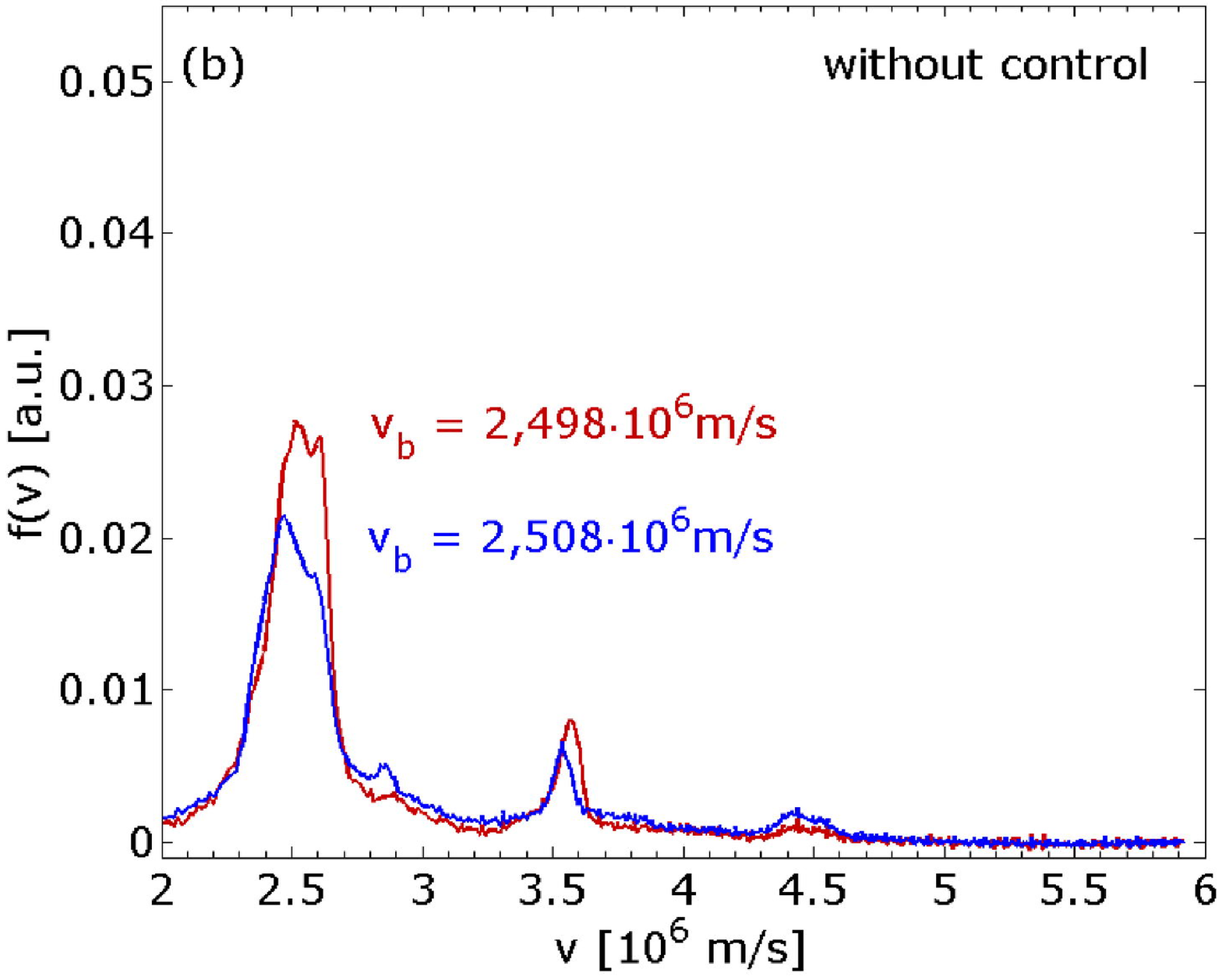}
\caption{Velocity distribution functions for two close values of
initial beam velocity (a) with and (b) without applied control .
The overlap parameter is $s=1.27$.}\label{fig:variazionevelo}
\end{figure}

Finally we check the sensitivity of the control mode with respect
to the initial beam velocity. This corresponds introducing an
error both on the wave number and on the amplitude of the control
mode. The overlap parameter $s$ depends on the phase velocity
difference between the helix and beam modes (see
Eq.(~\ref{eqn:pCh})); for such a reason we also measure the
non-controlled velocity distribution function for each initial
beam velocity. Figure~\ref{fig:varvelo}a clearly exhibits the
resonant condition expected at the reference value $2.51\times
10^6${\,\rm m/s}. We also note that, without control, chaos is
continuously increasing as expected since when the phase velocity
difference decreases resonance overlap (and chaos) increases.
Figure~\ref{fig:variazionevelo} shows how two beams with close
initial velocities with similar chaotic behavior have two
different responses to the same control term.

\vspace{1cm}

\section{Summary and conclusion}
\label{conclusions}

Even if a modification of the perturbation of a Hamiltonian system
generically leads to the enhancement of the chaotic behavior, we
have applied numerically and experimentally a general strategy and
an explicit algorithm to design a small but apt modification of
the potential which drastically reduces chaos and its attendant
diffusion by channeling chaotic transport. The experimental
results show that the method is tractable and robust, therefore
constituting a natural way to control the dynamics. The robustness
of the method has been checked for an overlap parameter equal to
$s=1.27$ by changing phase and amplitude of the control term and
beam velocity to check resonance condition on the helix dispersion
relation. All these measurements have shown a significant region
around the prescribed values for which the control is efficient.
The implementation is realized with an additional cost of energy
which corresponds to less than $1\%$ of the initial energy of the
two-wave system. We stress the importance of a fine tuning of the
parameters of the theoretically computed control term (e.g.,
amplitude, phase velocity) in order to force the experiment to
operate in a more regular regime. For such a reason an iterative
process to find some optimal experimental conditions is suggested
for future improvement of the method. Other control terms can be
used to increase stability (by taking into account the other
Fourier modes of $f$ given in Eq.~(\ref{eqn:fTWT}) when
experimentally feasible). The achievement of control and all the
tests on a TWT assert the possibility to practically control a
wide range of systems at a low additional cost of energy.
\vspace{1cm}
\section{Acknowledgment}
A.M. and F.D. are grateful to J-C.~Chezeaux, D.~Guyomarc'h, and
B.~Squizzaro for their skillful technical assistance, and to
D.F.~Escande and Y.~Elskens for fruitful discussions and a
critical reading of the manuscript. A. M. benefits from a grant by
Minist\`ere de la Recherche. This work is partially supported by
Euratom/CEA (contract EUR 344-88-1 FUA F).

\end{document}